\documentclass[twocolumn,showpacs,preprintnumbers,amsmath,amssymb]{revtex4}

\usepackage{graphicx}
\usepackage{dcolumn}
\usepackage{bm}

\begin{document}

\title{Randomizing world trade. I. A binary network analysis}

\author{Tiziano Squartini}
\affiliation{CSC and Department of Physics, University of Siena, Via Roma 56, 53100 Siena (Italy)}
\affiliation{Lorentz Institute for Theoretical Physics, Leiden Institute of Physics, University of Leiden, Niels Bohrweg 2, 2333 CA Leiden (The Netherlands)}
\author{Giorgio Fagiolo}
\affiliation{LEM, Sant'Anna School of Advanced Studies, 56127 Pisa (Italy)}
\author{Diego Garlaschelli}
\affiliation{Lorentz Institute for Theoretical Physics, Leiden Institute of Physics, University of Leiden, Niels Bohrweg 2, 2333 CA Leiden (The Netherlands)}

\date{\today}

\begin{abstract}
The international trade network (ITN) has received renewed multidisciplinary interest due to recent advances in network theory. However, it is still unclear whether a network approach conveys additional, nontrivial information with respect to traditional international-economics analyses that describe world trade only in terms of local (first-order) properties. In this and in a companion paper, we employ a recently proposed randomization method to assess in detail the role that local properties have in shaping higher-order patterns of the ITN in all its possible representations (binary/weighted, directed/undirected, aggregated/disaggregated by commodity) and across several years. Here we show that, remarkably, the properties of all binary projections of the network can be completely traced back to the degree sequence, which is therefore maximally informative. Our results imply that explaining the observed degree sequence of the ITN, which has not received particular attention in economic theory, should instead become one the main focuses of models of trade. 
\end{abstract}



\maketitle

\section{Introduction}
The network of import/export trade relationships among all world countries, known in the literature as the International Trade Network (ITN) or the World Trade Web (WTW), has received a renewed multidisciplinary interest in recent years \cite{LiC03,SeBo03,Garla2004,Garla2005,serrc07,Bhatta2007a,Bhatta2007b, Garla2007,Tzekina2008,Fagiolo2008physa,Fagiolo2008acs,Fagiolo2009pre,BariFagiGarla2010}, due to impressive advances in both empirical and theoretical approaches to the study of complex networks \cite{guidosbook, AlbertBarabasi2002, DoroMendes2003}. A number of robust patterns in the structure of this network have been empirically observed, both in its binary (when only the presence of a trade interaction is considered, irrespective of its intensity) and weighted (when also the magnitude of trade flows is taken into account) description. These stylized facts include local properties as well as higher-order patterns. Local properties involve direct (first-order) interactions alone, resulting in simple quantities such as node degree (the number of trade partners of a country), node strength (total trade volume of a country), and their directed-network analogues (i.e., when these statistics are computed taking into account edge/trade directionality). Higher-order characteristics are more complicated structural properties that also involve indirect interactions, i.e. topological paths connecting a country to the neighbors of its neighbors, or to countries farther apart. Examples include degree-degree correlations, average nearest-neighbor indicators, and clustering coefficients, to name just a few of them.

In general, local and higher-order topological properties are not independent of each other. In particular, even if one assumes that the network is formed as the result of local constraints alone, with higher-order properties being only the mere outcome of a specification of these constraints, it turns out that so-called structural correlations are automatically generated. Structural correlations sometimes appear as complicated patterns that might be confused with genuine correlations involving higher-order statistics, and interpreted as the presence of an additional level of topological organization. Therefore, in any real network it is important to characterize structural correlations and filter them out in order to assess whether nontrivial effects due to indirect interactions are indeed present.

In the specific case of the ITN, this problem is particularly important to assess whether the network formalism is really conveying additional, nontrivial information with respect to traditional international-economics analyses, which instead explain the empirical properties of trade in terms of country-specific macroeconomic variables alone. Indeed, the standard economic approach to the empirics of international trade \cite{Feenstra2004} has traditionally focused its analyses on the statistical properties of country-specific indicators like total trade, trade openness (ratio of total trade to GDP, i.e. Gross Domestic Product), number of trade partners, etc., that can be easily mapped to what, in the jargon of network analysis, one denotes as local properties or first-order node characteristics. Ultimately, understanding whether network analyses go a step beyond with respect to standard trade theory amounts to assess the effects of indirect interactions in the world trade system. Indeed, a wealth of results about the analysis of international trade have already been derived in the macroeconomics literature \cite{Feenstra2004} without making explicit use of the network description, and focusing on the above country-specific quantities alone. Whether more recent analyses of trade, directly inspired by the network paradigm
\cite{LiC03,SeBo03,Garla2004,Garla2005,serrc07,Bhatta2007a,Bhatta2007b,Garla2007,Tzekina2008,Fagiolo2008physa,Fagiolo2008acs,Fagiolo2009pre}, are indeed conveying additional and nontrivial information about the structure of international import/export flows, crucially depends on the answer to the above question. Some network-inspired studies have already tried to address this problem, but with ambiguous results. In some cases, it was suggested that local properties are enough to explain higher-order patterns \cite{Garla2004,mylikelihood,Bhatta2007a}, while in others the opposite conclusion was reached \cite{multipolarization}. However, previous analyses of the ITN focused on heterogeneous representations (either binary \cite{SeBo03,Garla2004} or weighted \cite{Bhatta2007a,Fagiolo2008physa,multipolarization}, either directed 	\cite{Garla2005,myreciprocity,serrc07} or undirected \cite{SeBo03,Garla2004}, either aggregated \cite{SeBo03,Garla2004,Fagiolo2008physa} or disaggregated \cite{BariFagiGarla2010} in separate commodities) and using different datasets, making consistent conclusions impossible.

In this and in a companion paper \cite{part2}, we explicitly address this problem and exploit a recently proposed analytical method \cite{myrandomization} to obtain, for any given topological property of interest, the value of the corresponding quantity averaged over the family of all randomized variants of the ITN that preserve the observed local properties. This allows us to identify empirical deviations from locally-induced structural correlations. Null models are used in our exercises to uncover significant features of the network and to understand to which extent some network statistics are sufficient to explain other network statistics. Our analysis is not however involved in explaining the underlying causal mechanisms shaping the network. Therefore, throughout this and its companion paper, we shall use the term ``explaining'' in a weak sense. For example, finding that a local network statistics X ``explains'' a higher-order network statistics Y in our null model will signal the presence of a strong correlation between the two statistics, so that X can be sufficient to fully reproduce Y in the network. Of course, we do not aim at using our null model to identify subtle causal links between X and Y, which in the real-world may be caused e.g. by some omitted variables that cause in a proper way the high observed correlation between X and Y.

In this first paper, we focus on the ITN as a binary network. We find that higher-order patterns of all binary (either directed or undirected) projections of the ITN are remarkably well explained by local properties alone (the degree sequences). This result is robust to different levels of commodity aggregation: even if with an increasing scatter, the degree sequence preserves its complete informativeness  as more disaggregated and sparser commodity-specific networks are considered. Moreover, we perform a temporal analysis and check the robustness of these results over time. Therefore we obtain, for the first time in this type of study, a detailed and homogeneous assessment of the role of local properties across different representations of the trade network, using various levels of commodity aggregation, and over several years.
From an international-trade perspective, our results indicate that binary network descriptions of trade can be significantly simplified by considering the degree sequence(s) only. In other words, in any binary representation of the ITN, the degree sequence turns out to be maximally informative, since its knowledge conveys almost the entire information about the topology of the network. 

In the companion paper \cite{part2}, we show that the picture changes completely when considering the ITN as a weighted network.
We find that the ITN is an excellent example of a network whose local topological properties cannot be deduced from its local weighted properties. 
These results highlight an important limitation of current economic models of trade, that do not aim at explaining or reproducing the observed degree sequence but focus more on the structure of weights \cite{GravityBook}. In other words, standard models of trade in economics have been focusing only on explaining the (positive) flow between any two countries, disregarding to a great extent theories that are able to account for the determinants of the creation of a link (i.e. the transition from a zero trade-flow to a positive trade flow). The observed extreme informativeness of the degree sequence leads us to conclude that such models should be substantially revised in order to  explicitly include the degree sequence of the ITN among the key properties to reproduce.

\section{Data and methods}
This Section describes the data we use to construct the various representations of the network in this and in the following paper \cite{part2}, discusses how the
country-specific properties that are usually considered in world-trade economics translate into local topological properties of the ITN, and discusses how these properties should be kept as constraints of our analysis using an appropriate network randomization method.

\subsection{The International Trade Network\label{sec_data}}
We use yearly bilateral data on exports and imports from the United Nations Commodity Trade Database (UN COMTRADE) \footnote{\texttt{http://comtrade.un.org/}.} from year 1992 to 2002. We have chosen this database because, despite its relatively short time interval (11 years), it contains trade data between countries disaggregated across commodity categories. This allows us to perform our analyses both at the aggregate level (total trade flows) and at the commodity-specific level, e.g. investigating whether local properties are sufficient to explain higher-order ones in commodity-specific networks of trade.

In order to perform a temporal analysis and allow comparisons across different years, we restrict ourselves to a balanced panel of $N=162$ countries that are present in the data throughout the time interval considered. 
As to the level of disaggregation, we choose the classification of trade values into $C=97$ possible commodities listed according to the Harmonized System 1996 (HS1996) \footnote{\texttt{http://unstats.un.org/}.}. Accordingly, for a given year $t$ we consider the trade value $e^c_{ij}(t)$ corresponding to exports of the particular commodity $c$ ($c=1,\dots C$).
Since, for every commodity, exports from country $i$ to country $j$ are reported twice (by both the importer and the exporter) and the two figures do not always match, we follow Ref.~\cite{BariFagiGarla2010} and only employ the flow as reported by the importer.
Besides commodity-specific data, we also compute the total value $e^0_{ij}(t)$ of exports from country $i$ to country $j$ as the sum over the exports of all $C=97$ commodity classes: 
\begin{equation}
e^0_{ij}(t)\equiv\sum_{c=1}^C e^c_{ij}(t)
\label{eq_e0}
\end{equation}
The particular aggregation procedure described above, which coincides with the one performed in Ref.~\cite{BariFagiGarla2010}, allows us to compare our analysis of the $C$ commodity-specific networks with a $(C+1)$-th aggregate network, avoiding possible inconsistencies between aggregated and disaggregated trade data. We stress that the resulting aggregated network data are in general different from those used in other analyses \cite{Garla2004,Garla2005,Fagiolo2009pre} of the same network. Nonetheless, as we show below, when we analyze network properties that have also been studied in previous studies of aggregate trade, we find perfect agreement.

\begin{table*}
\center \small
\begin{tabular}{cp{6cm}ccc}
\hline
\hline
   HS Code &  Commodity & Value (USD) & Value per link (USD) & $\%$ of aggregate trade \\
\hline
        84 & Nuclear reactors, boilers, machinery and mechanical appliances; parts thereof  &   $5.67\times 10^{11}$ &   $6.17\times 10^{7}$ &    11.37\% \\

        85 & Electric machinery, equipment and parts; sound equipment; television equipment  &   $5.58\times 10^{11}$ &   $6.37\times 10^{7}$ &    11.18\% \\

        27 & Mineral fuels, mineral oils \& products of their distillation; bitumin substances; mineral wax  &   $4.45\times 10^{11}$ &   $9.91\times 10^{7}$ &     8.92\% \\

        87 & Vehicles, (not railway, tramway, rolling stock); parts and accessories  &   $3.09\times 10^{11}$ &   $4.76\times 10^{7}$ &     6.19\% \\

        90 & Optical, photographic, cinematographic, measuring, checking, precision, medical or surgical instruments/apparatus; parts \& accessories  &   $1.78\times 10^{11}$ &   $2.48\times 10^{7}$ &     3.58\% \\

        39 & Plastics and articles thereof.  &   $1.71\times 10^{11}$ &   $2.33\times 10^{7}$ &     3.44\% \\

        29 & Organic chemicals  &   $1.67\times 10^{11}$ &   $3.29\times 10^{7}$ &     3.35\% \\

        30 & Pharmaceutical products  &    $1.4\times 10^{11}$ &   $2.59\times 10^{7}$ &     2.81\% \\

        72 & Iron and steel  &   $1.35\times 10^{11}$ &   $2.77\times 10^{7}$ &     2.70\% \\

        71 & Pearls, precious stones, metals, coins, etc &   $1.01\times 10^{11}$ &   $2.41\times 10^{7}$ &     2.02\% \\

        10 &   Cereals  &   $3.63\times 10^{10}$ &   $1.28\times 10^{7}$ &     0.73\% \\

        52 & Cotton, including yarn and woven fabric thereof  &   $3.29\times 10^{10}$ &   $6.96\times 10^{6}$ &     0.66\% \\

         9 & Coffee, tea, mate \& spices  &   $1.28\times 10^{10}$ &   $2.56\times 10^{6}$ &     0.26\% \\

        93 & Arms and ammunition, parts and accessories thereof &   $4.31\times 10^{9}$ &   $2.46\times 10^{6}$ &     0.09\% \\

       ALL &  Aggregate (all 97 commodities) &   $4.99\times 10^{12}$ &   $3.54\times 10^{8}$ &   100.00\% \\
       \hline
\hline
\end{tabular}
\caption{The 14 most relevant commodity classes (plus aggregate trade) in year 2003 and the corresponding 
total trade value (USD), trade value per link (USD), and share of world
aggregate trade. From Ref.~\cite{BariFagiGarla2010}.} \label{table}
\end{table*}

The quantities $\{e^c_{ij}(t)\}$ (where $c=0,\dots C$) defined above are the fundamental data that allow us to obtain different possible representations of the trade network, as well as the corresponding randomized counterparts (see below for the units of measure we adopted).
When we regard the ITN as a weighted directed network, we define the weight of the link from country $i$ to country $j$ in year $t$ for commodity $c$ as
\begin{equation}
w^c_{ij}(t)\equiv \lfloor e^c_{ij}(t) \rceil\qquad c=0,\dots C
\label{eq_wdir}
\end{equation}
where $\lfloor x \rceil\in\mathbb{N}$ denotes the nearest integer to the non-negative real number $x$. 
When we adopt a weighted but undirected (symmetrized) description, we define the weight of the link between countries $i$ and $j$ in year $t$ for commodity $c$ as
\begin{equation}
w^c_{ij}(t)\equiv w^c_{ji}(t)\equiv\left\lfloor \frac{e^c_{ij}(t)+e^c_{ji}(t)}{2}\right\rceil\qquad c=0,\dots C
\label{eq_wund}
\end{equation}
Therefore, in both the directed and undirected case, $w^c_{ij}(t)$ is an integer quantity.
Since in both cases we shall be interested in tracking the temporal evolution of most quantities, we also define rescaled weights (relative to the total yearly trade flow) as follows:
\begin{equation}
\tilde{w}^c_{ij}(t)\equiv\frac{w^c_{ij}(t)}{w^c_{tot}(t)}\qquad c=0,\dots C
\label{eq_tilde}
\end{equation}
where in the directed case $w^c_{ij}(t)$ is given by Eq.~(\ref{eq_wdir}) and $w^c_{tot}(t)\equiv\sum_i\sum_{j\ne i}w^c_{ij}(t)$ (the double sum runs over all $N(N-1)$ ordered pairs of vertices), while in the undirected case $w^c_{ij}(t)$ is given by Eq.~(\ref{eq_wund}) and $w^c_{tot}(t)\equiv\sum_i\sum_{j< i}w^c_{ij}(t)$ (the double sum runs over all the $N(N-1)/2$ unordered pairs).
In such a way, trend effects are washed away and we obtain adimensional weights that are automatically deflated, allowing consistent comparisons across different years and different commodities.

In the binary representations of the network, we draw a link from $i$ to $j$ whenever the corresponding weight $w^c_{ij}$ is strictly positive. If $\Theta(x)$ denotes the step function (equal to $1$ if $x>0$ and $0$ otherwise), the adjacency matrix of the binary projection of the network in year $t$ for commodity $c$ is
\begin{equation}
a^c_{ij}(t)\equiv \Theta[w^c_{ij}(t)]\qquad c=0,\dots C
\label{eq_a}
\end{equation}
where $w^c_{ij}(t)$ is given either by Eq.~(\ref{eq_wdir}) or by Eq.~(\ref{eq_wund}), depending on whether one is interested in a directed or undirected binary projection of the network respectively. 

For each of the $C+1$ commodity categories, we can consider four network representations (binary undirected, binary directed, weighted undirected, weighted directed).
When reporting our results, we will first describe the aggregated networks ($c=0$) and then the disaggregated (commodity-specific) ones. In particular, among the 97 commodity classes, we will focus on the 14 particularly relevant commodities identified in Ref.~\cite{BariFagiGarla2010}, which are reported in table~\ref{table}. These 14 commodities include the 10 most traded commodities ($c=84, 85, 27, 87, 90, 39, 29, 30, 72, 71$ according to the HS1996) in terms of total trade value (following the ranking in year 2003 \cite{BariFagiGarla2010}), plus 4 classes ($c=10, 52, 9, 93$ according to the HS1996) which are less traded but more relevant in economic terms.
Taken together, the 10 most traded commodities account for $56\%$ of total world trade in 2003; moreover, they also feature the largest values of trade value per link (also shown in the table). The 14 commodities considered account together for $57\%$ of world trade in 2003. As an intermediate level of aggregation, we shall also consider the networks formed by the sum of these 14 commodities. The original data $\{e^{c}_{ij}(t)\}$ are available in current U.S. dollars (USD) for all commodities; however, due to the different trade volumes involved, we use different units of measure for different levels of aggregation \footnote{We left $\{e^{9}_{ij}(t)\}$ and $\{e^{93}_{ij}(t)\}$ in current U.S. dollars (USD); we divided $\{e^{39}_{ij}(t)\}$ and $\{e^{90}_{ij}(t)\}$ by 10; we divided $\{e^{84}_{ij}(t)\}$ and the sum of the top 14 commodities by 100; we divided aggregate data ($\{e^{0}_{ij}(t)\}$) by 1000. Accordingly, after the rounding defined by eqs.(\ref{eq_wdir}) and (\ref{eq_wund}), we obtained trade flows $\{w^{c}_{ij}(t)\}$ expressed in integer multiples of either 1 USD, 10 USD, 100 USD, or 1000 USD.}.

\subsection{Controlling for local properties\label{sec_controlling}}
As we mentioned, our main interest in the present work is assessing whether higher-order properties of the ITN can be simply traced back to local properties, which are the main focus of traditional macroeconomic analyses of international trade. Such standard country-specific properties include: total exports, total imports, total trade (sum of total exports and total imports), trade openness (ratio of total trade to GDP), the number of countries whom a country exports to and imports from, the total number of trade partners (irrespective of whether they are importers or exporters, or both). All these quantities can be simply obtained as local sums over direct interactions (countries one step apart) in a suitable representation of the network. 

For instance, the number of trade partners of country $i$ is simply the number of neighbors of node $i$ in the binary undirected projection, i.e. the \emph{degree}
\begin{equation}
k_i\equiv \sum_{j\ne i} a_{ij}
\label{eq_controlk}
\end{equation}
In the above equation and in what follows, we drop the dependence of topological quantities on the particular year $t$ for simplicity. We also drop the superscript $c$ specifying a particular commodity, as all the formulas hold for any $c$. This means that, if the aggregated network of total trade is considered, then $a_{ij}$ and $w_{ij}$ represent the aggregate quantities $a^0_{ij}$ and $w^0_{ij}$, where the commodity $c=0$ formally represents the sum over all commodities, as in Eq.~(\ref{eq_e0}). Otherwise, if the commodity-specific network involving only the trade of the particular commodity $c$ (with $c>0$) is considered, then $a_{ij}$ and $w_{ij}$ represent the values $a^c_{ij}$ and $w^c_{ij}$ for that commodity.

The number of countries whom a country exports to and imports from are simply the two directed analogues (the \emph{out-degree} $k^{out}_i$ and the \emph{in-degree} $k^{in}_i$ respectively) of the above quantity in the binary directed description:
\begin{eqnarray}
k^{out}_i&\equiv& \sum_{j\ne i} a_{ij}\label{eq_controlkout}\\
k^{in}_i&\equiv& \sum_{j\ne i} a_{ji}\label{eq_controlkin}
\end{eqnarray}

Similarly, as evident from Eq.~(\ref{eq_wund}), country $i$'s total trade coincides with twice the sum of weights reaching node $i$ in the weighted undirected representation, i.e. the \emph{strength}
\begin{equation}
s_i\equiv \sum_{j\ne i} w_{ij}
\label{eq_controls}
\end{equation}

Finally, total exports (imports) of country $i$ are simply the sum of out-going (in-coming) weights in the weighted directed representation of the ITN. These quantities are known as the \emph{out-strength} $s^{out}_i$ and \emph{in-strength} $s^{in}_i$ of node $i$:
\begin{eqnarray}
s^{out}_i&\equiv& \sum_{j\ne i} w_{ij}\label{eq_controlsout}\\
s^{in}_i&\equiv& \sum_{j\ne i} w_{ji}\label{eq_controlsin}
\end{eqnarray}

Another country-specific property which is widely used as an explanatory variable of trade patterns is the GDP or the \emph{per capita} GDP (i.e. the ratio of GDP to population). This property is sometimes used to rescale trade values, as in the case of trade openness which is defined as a country's ratio of total trade to GDP. Unlike the quantities discussed above, the GDP is not a topological entity. Nonetheless, it is empirically observed to be positively (and strongly) correlated with the degree \cite{Garla2004} and with node strength \cite{Fagiolo2009pre} (we will comment more on this in Section \ref{sec_bun}). Therefore, even if this is not the main aim of the present work, one should be aware that assessing the role of local topological properties also indirectly implies, to a large extent, assessing the role of the GDP of countries.

\subsection{Rewiring the ITN}
We showed that, in a network language, the standard country-specific properties used to characterize world trade translate into simple local topological properties of the ITN. This naturally implies that, in our analysis, it is important to consider a null model of the ITN where such properties are enforced as constraints, and the topology is otherwise maximally random. Different methods that produce randomized ensembles of networks with given constraints exist \cite{Holland_Leinhardt_1976,Kannan_etal_1999,Katz_Powell_1957,Rao_etal_1996,Roberts_2000,Shen-Orr_etal_2002,Snijders_1991,maslov,msz,chunglu,myrandomization}. As we mentioned, we aim at studying many topological properties of several different representations and temporal snapshots of the ITN. Therefore, we need a fast method that can deal with many networks in a relatively short time, and treat binary, weighted, directed and undirected graphs in a  consistent fashion. To this end, we employ the maximum-likelihood method introduced in ref. \cite{myrandomization}, which provides the expectation values (over the randomized ensemble) of the desired topological properties analytically, in contrast with alternative methods \cite{maslov,msz} which require to explicitly generate many randomized variants of the real network computationally.
Moreover, the method is density-independent and works for both sparse and dense networks. By contrast, other (analytical or computational) approaches are density-dependent and not optimized for dense networks: the Chung-Lu (analytical) approach \cite{chunglu} works only for sparse networks, and the Maslov-Sneppen (computational) algorithm \cite{maslov,msz} becomes too time consuming for dense networks. Since the ITN is an unusually dense network, the maximum-likelihood method is the natural choice that allows us to perform a detailed analysis, covering all possible representations across several years, which would otherwise require an impressive amount of time.

In the Appendix we describe the maximum-likelihood method in some detail, in particular its application to the topological properties of interest for the present case study.
Given any topological property $X$, the method provides the average value $\langle X\rangle$ of $X$ across the ensemble of random graphs with the same average (across the ensemble itself) constraints as the real network. For simplicity, in this and in the companion paper we sometimes denote $\langle X\rangle$ as a \emph{randomized property}, and its value as the \emph{randomized value} of $X$, even if technically no randomization process has been required (all the results have been obtained analytically).
Similarly, we imagine the graph ensemble as a  \emph{rewired} version of the original network, even if no rewiring has taken place explicitly.

\section{The ITN as a binary undirected network\label{sec_bun}}
As we mentioned in Section \ref{sec_data}, in its binary representation the ITN is defined as a graph whose edges report the presence of trade relationships among world countries, irrespective of the intensity of these relationships. The binary representation of the ITN can be either undirected or directed, depending on whether one is interested in specifying the orientation of trade flows. In both cases, the complete information about the topology of the network is encoded in the adjacency matrix $\mathbf{A}$, whose entries $\{a_{ij}\}$ are defined as in Eq.~(\ref{eq_a}).

In the simplest case, the presence of at least one of the two possible trade relationships between any two countries $i$ and $j$ (either from $i$ to $j$ or from $j$ to $i$) is represented as one undirected edge between nodes $i$ and $j$. Therefore $a_{ij}=a_{ji}$ and $\mathbf{A}$ is a symmetric matrix. In this binary undirected description, as shown in Eq.~(\ref{eq_controlk}), the local constraints $\{C_a\}$ are the degrees of all vertices, i.e. the \emph{degree sequence} $\{k_i\}$. 
Therefore, the maximum-likelihood randomization method \cite{myrandomization} (see Appendix) works by specifying the constraints $\{C_a\}\equiv \{k_i\}$ and allows us to write down the probability of any graph $\mathbf{G}$ in the grandcanonical ensemble, which is uniquely specified by its generic adjacency matrix $\mathbf{A}$. As summarized in Appendix \ref{app_bun}, this allows us to easily obtain the expectation value $\langle X\rangle$, formally defined in Eq.~(\ref{eq_X}), of any property $X$ across the ensemble of binary undirected graphs whose expected degree sequence is equal to the empirical one.
Note that, among the possible properties, the degree of vertices plays a special role, as its expectation value $\langle k_i\rangle$ is exactly equal to the empirical value $k_i$, as required by the method. Therefore the values $\{k_i\}$ are useful control parameters and can be efficiently  used as independent variables in terms of which other properties $X$ can be visualized.

For the sake of simplicity, in Sections \ref{sec_bun_annd} and \ref{sec_bun_clustering} we first report the results of this analysis on a single snapshot of the commodity-aggregated network (the last year in our temporal window, i.e. 2002). Then, we discuss the robustness of our results through time by tracking them backwards in Section \ref{sec_bun_evolution}. We finally consider the disaggregated analysis of commodity-specific networks in Section \ref{sec_bun_dis}.
\begin{figure}[t]
\includegraphics[width=.45\textwidth]{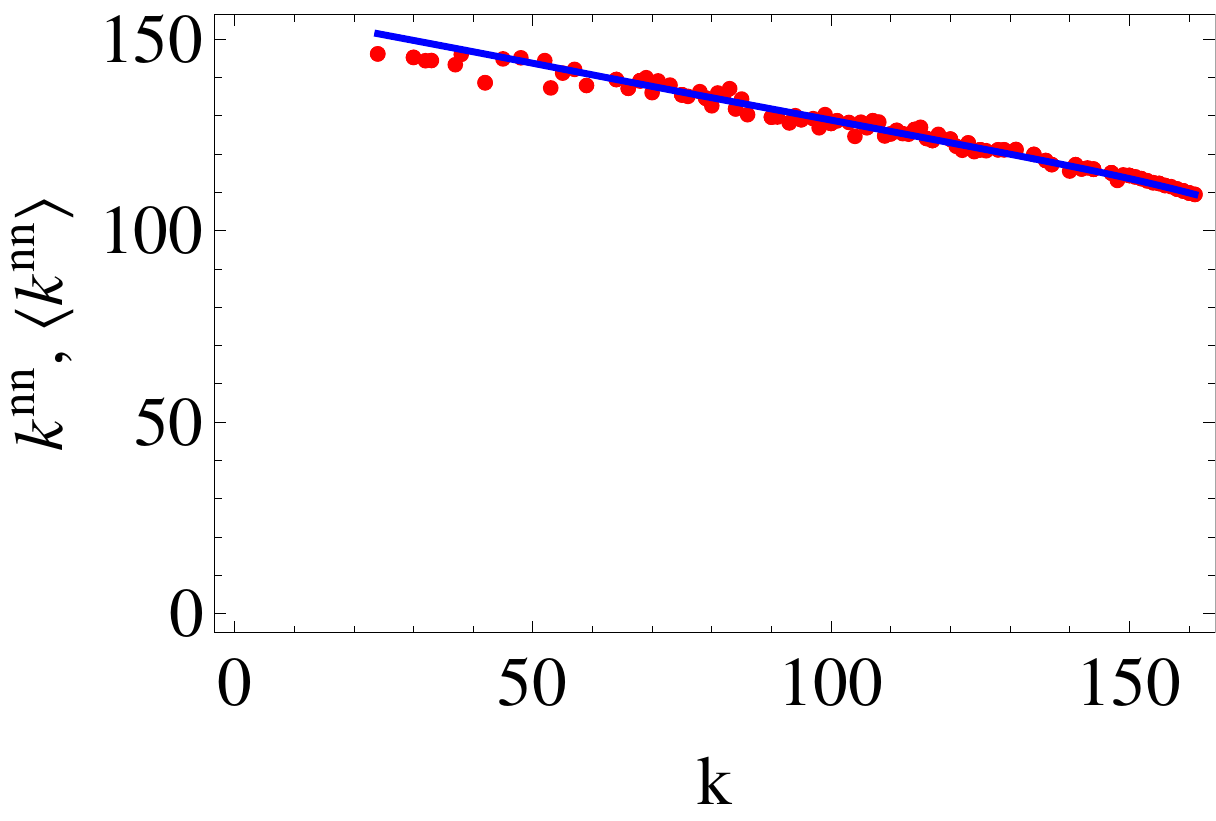}
\caption{\label{fig_bun_knn} (Color online) Average nearest neighbor degree $k^{nn}_i$ versus degree $k_i$ in the 2002 snapshot of the real binary undirected ITN (red points), and corresponding average over the maximum-entropy ensemble with specified degrees (blue solid curve).}
\end{figure}

\subsection{Average nearest neighbor degree\label{sec_bun_annd}}
We start with the analysis of the aggregated version of the ITN, representing the trade of all commodities ($c=0$ in our notation). In the following formulas, the matrix $\mathbf{A}$ therefore denotes the aggregate matrix $\mathbf{A}^0$, where we drop the superscript for brevity.
As a first quantity, we consider the  \emph{average nearest neighbor degree} (ANND) of vertex $i$, defined as
\begin{equation}
k^{nn}_i\equiv\frac{\sum_{j\ne i}a_{ij}k_j}{k_i}
=\frac{\sum_{j\ne i}\sum_{k\ne j}a_{ij}a_{jk}}{\sum_{j\ne i}a_{ij}}
\label{eq_bun_knn}
\end{equation}
and measuring the average number of partners of the neighbors of a given node $i$. The above quantity involves indirect interactions of length two, as evidenced from the presence of terms of the type $a_{ij}a_{jk}$ in the definition.
Whether these $2$-paths are a simple outcome of the concatenation of two independent edges can be inspected by considering the correlation structure of the network, and in particular by plotting $k^{nn}_i$ versus $k_i$. The result is shown in Fig.~\ref{fig_bun_knn}. We observe a decreasing trend, confirming what already found in previous studies employing different datasets \cite{SeBo03, Garla2005, Fagiolo2009pre}. This means that countries trading with highly connected countries have a few trade partners, whereas countries trading with poorly connected countries have many trade partners. This correlation profile, known as \emph{disassortativity}, might signal an interesting pattern in the trade network. However, if we compare this trend with the one followed by the corresponding randomized quantity $\langle k^{nn}_i\rangle$ (see Appendix \ref{app_bun} for its expression), we find that the two behaviors coincide. 
This is an important effect of structural constraints in a dense network \cite{newman_origin}: contrary to what naively expected \cite{pastorvespignani}, even in a network where links are drawn randomly between vertices with given heterogeneous degrees, the ANND is not constant. This means that the degree sequence constrains the correlation structure, and that it is impossible to have a flat profile ($k^{nn}_i$ independent of $k_i$) unless one forces the system to display it by introducing additional mechanisms (hence additional correlations of opposite sign).
\begin{figure}[t]
\includegraphics[width=.45\textwidth]{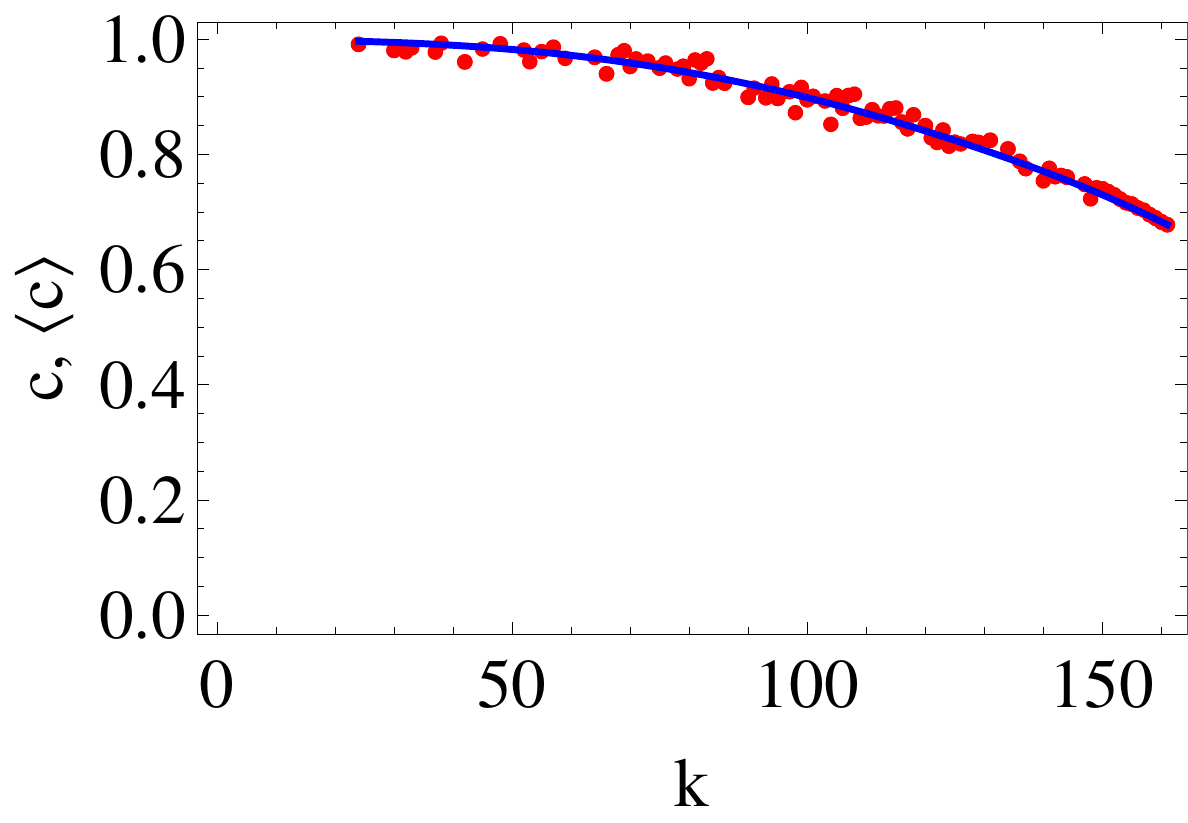}
\caption{\label{fig_bun_ck} (Color online) Clustering coefficient $c_i$ versus degree $k_i$ in the 2002 snapshot of the real binary undirected ITN (red points), and corresponding average over the maximum-entropy ensemble with specified degrees (blue solid curve).}
\end{figure}

\subsection{Clustering coefficient\label{sec_bun_clustering}}
A similar result is found for the behavior of the \emph{clustering coefficient} $c_i$, representing the fraction of pairs of neighbors of vertex $i$ which are also neighbors of each other:
\begin{eqnarray}
c_i&\equiv&\frac{\sum_{j\neq i}\sum_{k\ne i,j}a_{ij}a_{jk}a_{ki}}{k_{i}(k_{i}-1)}\nonumber\\
&=&\frac{\sum_{j\neq i}\sum_{k\ne i,j}a_{ij}a_{jk}a_{ki}}{\sum_{j\neq i}\sum_{k\ne i,j}a_{ij}a_{ik}}
\label{eq_bun_ck}
\end{eqnarray}
The clustering coefficient is a measure of the fraction of potential triangles attached to $i$ that are actually realized. This means that indirect interactions of length three, corresponding to products of the type $a_{ij}a_{jk}a_{ki}$ entering Eq.~(\ref{eq_bun_ck}), now come into play.
Again, we find a decreasing trend of $c_i$ as a function of $k_i$ (see Fig.~\ref{fig_bun_ck}). This means that trade partners of highly connected countries are poorly interconnected, whereas partners of poorly connected countries are highly interconnected. However, if this trend is compared with the one displayed by the randomized quantity $\langle c_i\rangle$ (see Appendix \ref{app_bun}), we again find a very close agreement. This signals that in the ITN also the profile of the clustering coefficient is completely explained by the constraint on the degree sequence, and does not imply the presence of meaningful indirect interactions on top of a concatenation of direct interactions alone.

The above results show that the patterns observed in the binary undirected description of the ITN do not require, besides the fact that different countries have specific numbers of trade partners, the presence of higher-order mechanisms as an additional explanation. On the other hand, the fact that the degrees alone are enough to explain higher-order network properties means that the degree sequence is an important structural pattern in its own. This highlights the importance of reproducing the observed degree sequence in models of trade. We will comment more about this point later on.

\subsection{Evolution of binary undirected properties\label{sec_bun_evolution}}
We now check the robustness of the previous results through time.
This amounts to perform the same analysis on each of the 11 years in our time window ranging from 1992 to 2002. For each of these snapshots, we specify the degree sequence and evaluate the maximally random ensemble of binary undirected graphs. We then compare each observed property $X$ with the corresponding average $\langle X\rangle$ (repeating the procedure described in Appendix \ref{app_bun}) over the null model for that specific year. We systematically find the same results described above for each and every snapshot. For visual purposes, rather than replicating the same plots shown above for all the years considered, we choose a more compact description of the observed patterns and portray its temporal evolution in a simple way. As we now show, this also provides us with a characterization of various temporal trends displayed by each topological property, conveying more information than a fixed-year description of the trade system.

\begin{figure}[t]
\includegraphics[width=.45\textwidth]{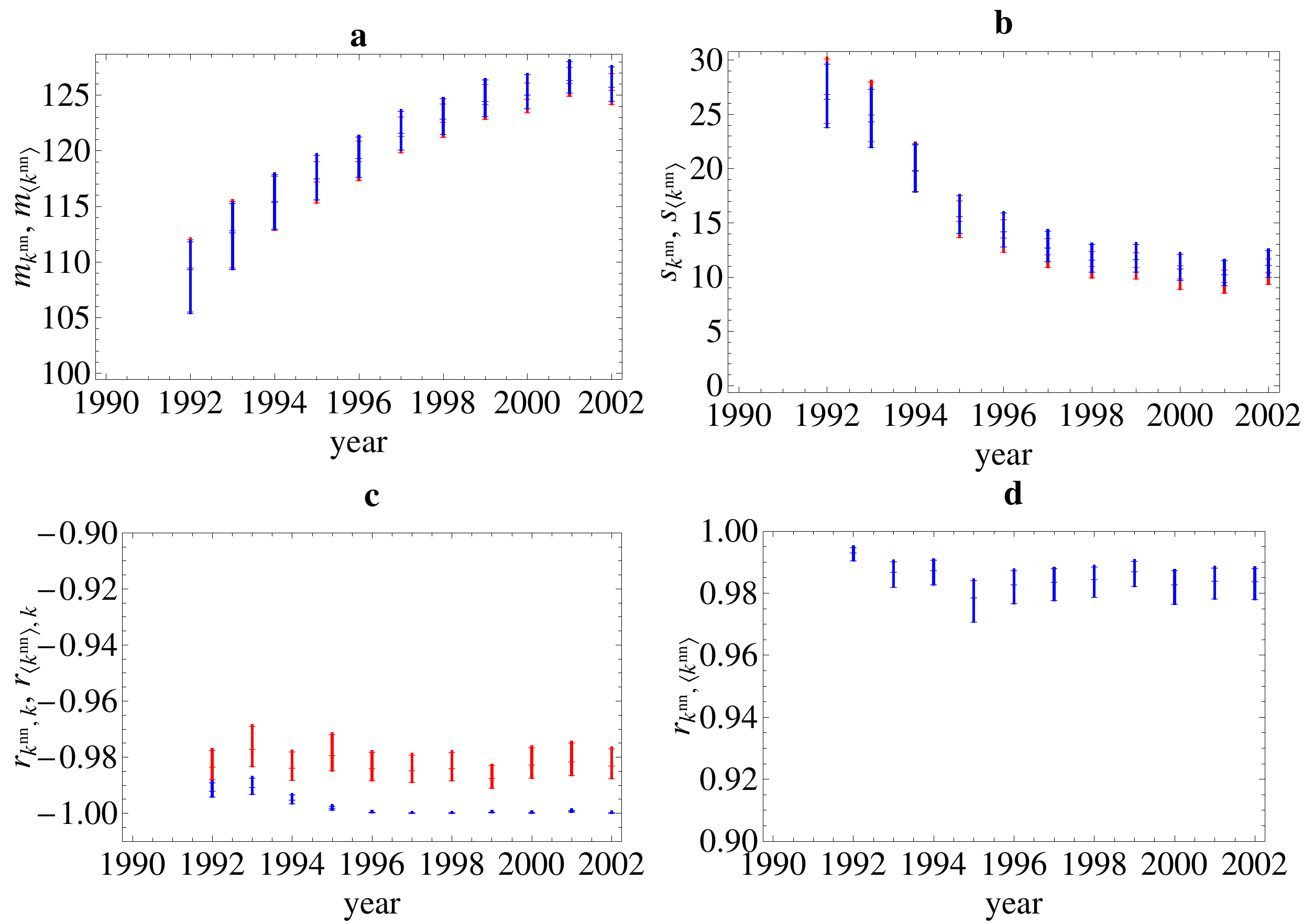}
\caption{\label{fig_bun_knn_t}
(Color online) Temporal evolution of the properties of the nearest neighbor degree $k^{nn}_i$ in the 1992-2002 snapshots of the real binary undirected ITN and of the corresponding maximum-entropy null model with specified degrees. \textbf{a)} average of $k^{nn}_i$ across all vertices (red: real data; blue: null model, indistinguishable from real data). \textbf{b)} standard deviation of $k^{nn}_i$ across all vertices (red: real data; blue: null model, indistinguishable from real data). \textbf{c)} correlation coefficient between $k^{nn}_i$ and $k_i$ (red, upper symbols: real data; blue, lower symbols: null model). \textbf{d)} correlation coefficient between $k^{nn}_i$ and $\langle k^{nn}_i\rangle$. The $95\%$ confidence intervals of all quantities are represented as vertical bars.}
\end{figure}

We first consider the average nearest neighbor degree. For a given year, we focus on the two lists of vertex-specific values $\{k^{nn}_i\}$ and $\{\langle k^{nn}_i\rangle\}$ for the real and randomized network respectively. We compute the average ($m_{k^{nn}}$ and $m_{\langle k^{nn}\rangle}$) and the associated $95\%$ confidence interval of both lists and plot them together as in Fig.~\ref{fig_bun_knn_t}a. We repeat this for all years and obtain a plot which informs us about the temporal evolution of the ANND in the real and randomized network separately. We find that the average value of the empirical ANND has been increasing steadily during the time period considered. However, the same is true for its randomized value, which is always consistent with the real one within the confidence intervals. This means that the null model completely reproduces the temporal trend of degree-degree correlations.

In principle, the increase of the ANND could be simply due to an overall increase in link density. To further study this possibility, we have compared the yearly growth rate ($X_t/X_{t-1}-1$) of the average ANND and of the link density in the period considered. We found two regimes: initially (from 1993 to 1997) the density has a larger (but decreasing) growth rate than the ANND, while from 1998 to 2002 onwards the two rates converge. Therefore it is useful to keep in mind that the evolution of the average ANND, as well as that of other average properties we consider below, is in general not merely reflecting the evolution of the overall link density.

In Fig.~\ref{fig_bun_knn_t}b we also plot the temporal evolution of the standard deviations $s_{k^{nn}}$ and $s_{\langle k^{nn}\rangle}$ (with associated $95\%$ confidence intervals) of the two lists of values $\{k^{nn}_i\}$ and $\{\langle k^{nn}_i\rangle\}$. We find that the variance of the empirical average nearest neighbor degree has been decreasing in time, but once more this behavior is completely reproduced by the null model and therefore fully explained by the evolution of the degree sequence alone.
Moreover, in Fig.~\ref{fig_bun_knn_t}c we show the Pearson (product-moment) correlation coefficient $r_{k^{nn},k}$ (with $95\%$ confidence interval) between $\{k^{nn}_i\}$ and $\{k_i\}$, and similarly the correlation coefficient $r_{\langle k^{nn}\rangle,k}$ between the randomized quantities $\{\langle k^{nn}_i\rangle\}$ and $\{k_i\}$ (recall that $\{\langle k_i\rangle\}=\{k_i\}$ by construction). This informs us in a compact way about the evolution of the dependence of the ANND on the degree, i.e. of the change in the structure of the scatter plot we showed previously in Fig.~\ref{fig_bun_knn}. We find that the disassortative character of the scatter plot results in a correlation coefficient close to $-1$, which has remained remarkably stable in time across the interval considered, and always very close to the randomized value.

The complete accordance between the real and randomized ANND in each and every snapshot is confirmed by Fig.~\ref{fig_bun_knn_t}d, where we show the correlation coefficient $r_{k^{nn},\langle k^{nn}\rangle}$ (with $95\%$ confidence interval) between the empirical ANND, $\{k^{nn}_i\}$, and the randomized one, $\{\langle k^{nn}_i\rangle\}$. We observe an approximately constant value close to $1$, signaling perfect correlation between the two quantities. This exhaustively explains the accordance between the real and randomized ANND for all vertices, while the other three panels of Fig.~\ref{fig_bun_knn_t} also inform about various overall temporal trends of the ANND, as we discussed.

\begin{figure}[t]
\includegraphics[width=.45\textwidth]{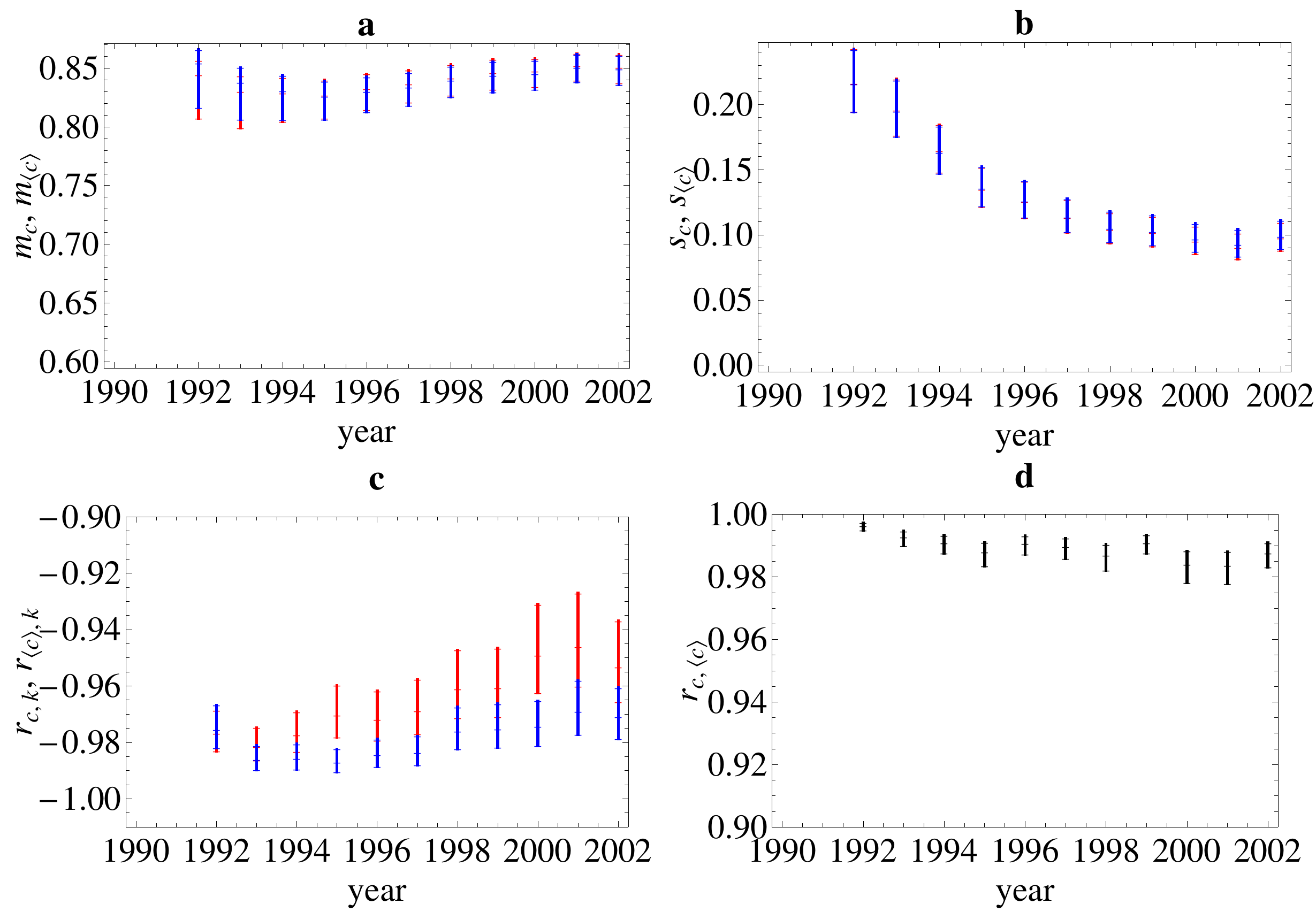}
\caption{\label{fig_bun_c_t}
(Color online) Temporal evolution of the properties of the clustering coefficient $c_i$ in the 1992-2002 snapshots of the real binary undirected ITN and of the corresponding maximum-entropy null model with specified degrees. \textbf{a)} average of $c_i$ across all vertices (red: real data; blue: null model, indistinguishable from real data). \textbf{b)} standard deviation of $c_i$ across all vertices (red: real data; blue: null model, indistinguishable from real data). \textbf{c)} correlation coefficient between $c_i$ and $k_i$ (red, upper symbols: real data; blue, lower symbols: null model). \textbf{d)} correlation coefficient between $c_i$ and $\langle c_i\rangle$. The $95\%$ confidence intervals of all quantities are represented as vertical bars.}
\end{figure}

In Fig.~\ref{fig_bun_c_t} we show the same analysis for the values $\{c_i\}$ and $\{\langle c_i\rangle\}$ of the clustering coefficient. In this case we observe an almost constant trend of the average clustering coefficient (Fig.~\ref{fig_bun_c_t}a), a decreasing standard deviation (Fig.~\ref{fig_bun_c_t}b), and a stable strong anticorrelation between clustering and degree (Fig.~\ref{fig_bun_c_t}c). Again, we find that the real and randomized values are always consistent with each other, so that the evolution of the empirical values is fully reproduced by the null model. This is confirmed by Fig.~\ref{fig_bun_c_t}d, which shows that the correlation between $\{c_i\}$ and $\{\langle c_i\rangle\}$ is always very close to $1$. As for the ANND, these results clearly indicate that the real and randomized values of the clustering coefficient of all vertices are always in perfect agreement, and that the temporal trends displayed by this quantity are completely explained by the evolution of the degree sequence.

\subsection{Commodity-specific binary undirected networks\label{sec_bun_dis}}
We complete our analysis of the ITN as a binary undirected network by studying whether the picture changes when one considers, rather than the network aggregating the trade of all types of commodities, the individual networks formed by imports and exports of single commodities. To this end, we focus on the disaggregated data described in Section \ref{sec_data} and we repeat the analysis reported above, by identifying the matrix $\mathbf{A}$ with various disaggregated matrices $\mathbf{A}^c$ (with $c>0$).

\begin{figure}[t]
\includegraphics[width=.45\textwidth]{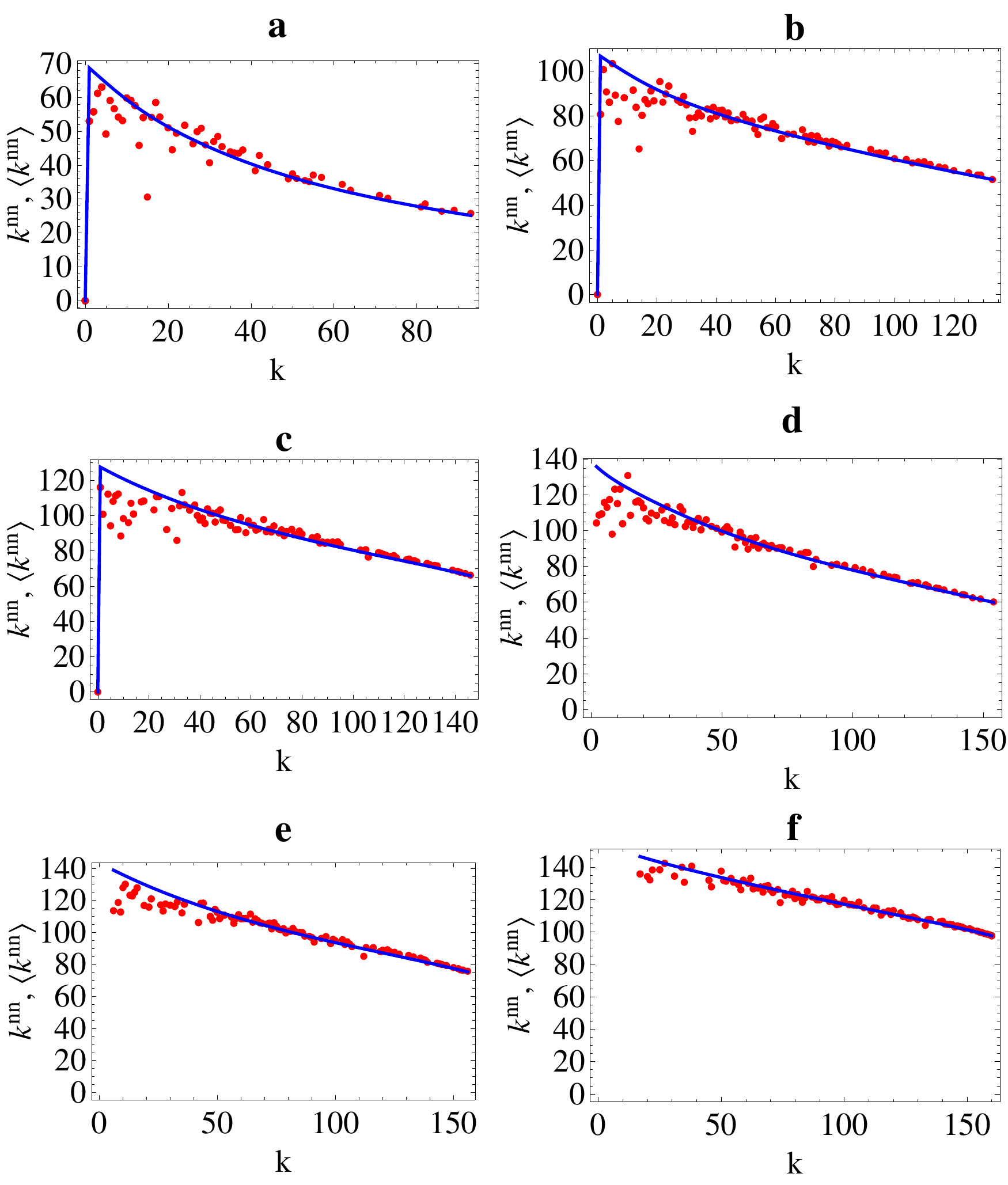}
\caption{\label{fig_bun_dis1}
(Color online) Average nearest neighbor degree $k^{nn}_i$ versus degree $k_i$ in the 2002 snapshots of the commodity-specific (disaggregated) versions of the real binary undirected ITN (red points), and corresponding average over the maximum-entropy ensemble with specified degrees (blue solid curves).
\textbf{a)} commodity 93; 
\textbf{b)} commodity 09; 
\textbf{c)} commodity 39; 
\textbf{d)} commodity 90; 
\textbf{e)} commodity 84; 
\textbf{f)} aggregation of the top 14 commodities (see Table \ref{table} for details). From a) to f), the intensity of trade and level of aggregation increases.} 
\end{figure}

\begin{figure}[t]
\includegraphics[width=.45\textwidth]{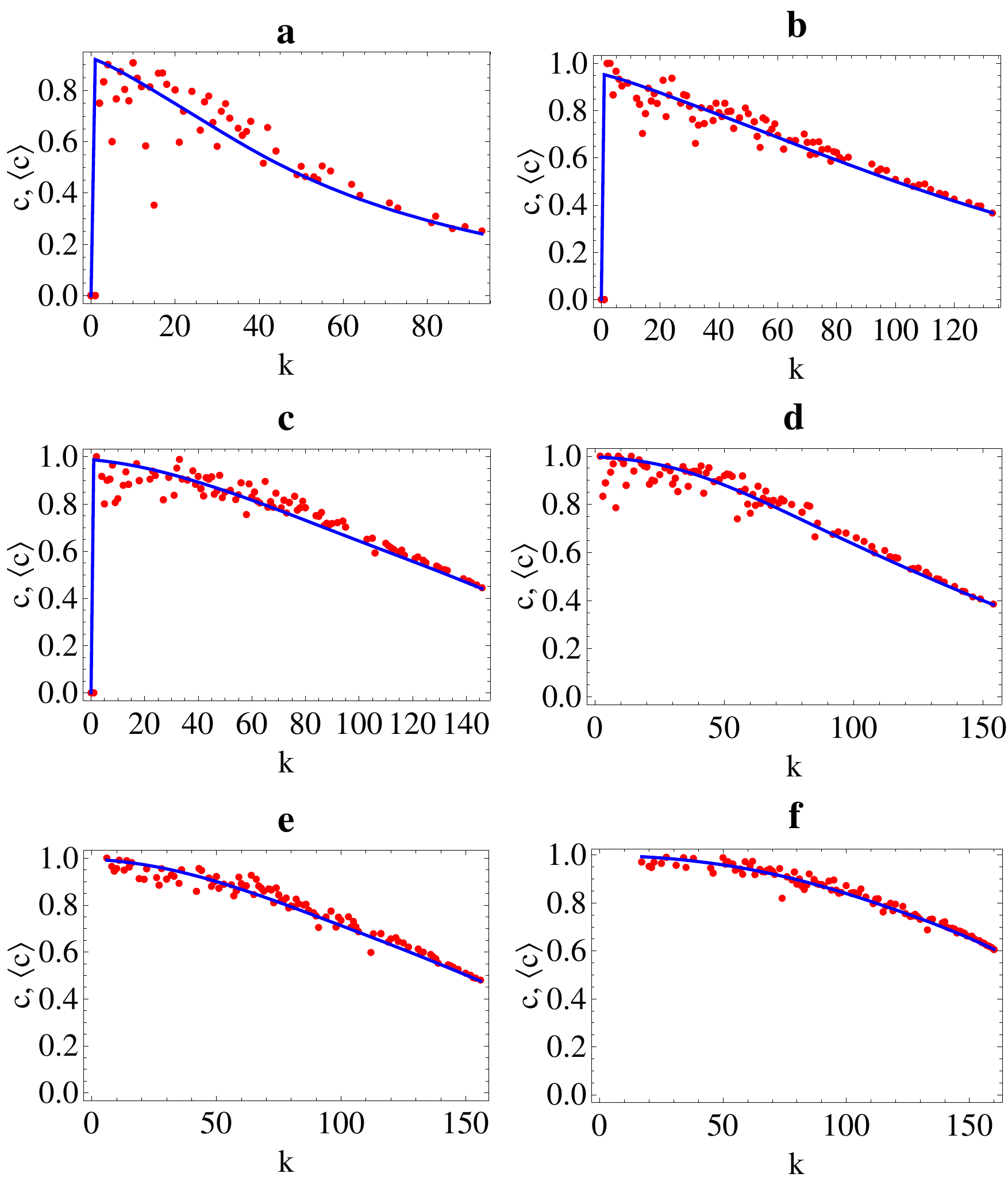}
\caption{\label{fig_bun_dis2}
(Color online) Clustering coefficient $c_i$ versus degree $k_i$ in the 2002 snapshots of the commodity-specific (disaggregated) versions of the real binary undirected ITN (red points), and corresponding average over the maximum-entropy ensemble with specified degrees (blue solid curves).
\textbf{a)} commodity 93; 
\textbf{b)} commodity 09; 
\textbf{c)} commodity 39; 
\textbf{d)} commodity 90; 
\textbf{e)} commodity 84; 
\textbf{f)} aggregation of the top 14 commodities (see Table \ref{table} for details).
From a) to f), the intensity of trade and level of aggregation increases.} 
\end{figure}

We find that the results obtained in our aggregated study also hold for individual commodities. For brevity, we only report the scatter plots of the average nearest neighbor degree (Fig.~ \ref{fig_bun_dis1}) and clustering coefficient (Fig.~ \ref{fig_bun_dis2}) for the 2002 snapshots of 6 commodity-specific networks. The 6 commodities are chosen among the top 14 reported in Table \ref{table}. In particular, we select the two least traded commodities in the set ($c=93,9$), two intermediate ones ($c=39,90$), the most traded one ($c=84$), plus the network formed by combining all the top 14 commodities, i.e. an intermediate level of aggregation between single commodities and the completely aggregated data ($c=0$), which we already considered in the previous analysis (Figs.~\ref{fig_bun_knn} and \ref{fig_bun_ck}). With the addition of the latter, the results shown span 7 different cases ordered by increasing trade intensity and level of commodity aggregation. Similar results hold also for the other commodities not shown.

If we compare Fig.~\ref{fig_bun_dis1} with Fig.~\ref{fig_bun_knn}, we see that the trend displayed by ANND in the aggregated network is preserved, even if with a slightly increasing scatter, as sparser and less disaggregated commodity classes are considered. Importantly, the accordance between real and randomized values is also preserved. The same is true for the clustering coefficient, cf. Fig.~\ref{fig_bun_dis2} and its comparison with Fig.~\ref{fig_bun_ck}. These results indicate that the degree sequence maintains its complete informativeness across different levels of commodity resolution, and irrespective of the corresponding intensity of trade.
Thus, remarkably, the knowledge of the number of trade partners involving only a specific commodity still allows to reproduce the properties of the corresponding commodity-specific network. 

As a summary of our binary undirected analysis we conclude that, in order to explain the evolution of the ANND and clustering of the ITN, it is unnecessary to invoke additional mechanisms besides those accounting for the evolution of the degree sequence alone. Since the ANND and clustering already probe the effects of indirect interactions of length two and three respectively, and since higher-order correlations involving longer topological paths are built on these lower-level ones, the null model we considered here is very likely to fully reproduce the properties of the ITN at all orders. In other words, we found that in the binary undirected representation of the ITN the degree sequence is maximally informative, as its knowledge allows to predict the higher-order topological properties of the network that we have explored in this and in the companion paper. An interesting question is whether the degree sequence is also able to reproduce other higher-order network properties, such as path lengths, node centrality, etc.. Whereas this study does not explicitly address this question, we argue that the answer will be positive in the light of the very nature of the ITN. Its high density indeed implies that path lengths are almost never larger than 3. As a consequence, network properties of order larger than 3 are typically well proxied by local properties. An example is the extremely high correlation between country centrality (measured e.g. in terms of betweenness centrality) and node degree, typically found in previous studies. The robustness of this result across several years and different commodity classes strengthens our previous discussion about the importance of including the degree sequence among the focuses of theories and models of trade, which are instead currently oriented mainly at reproducing the weighted structure, rather than the topology of the ITN.

\section{The ITN as a binary directed network}
We now consider the binary directed description of the ITN, with an interest in understanding whether the introduction of directionality changes the picture we have described so far. In the directed binary case, a graph $\mathbf{G}$ is completely specified by its adjacency matrix $\mathbf{A}$ which is in general not symmetric, and whose entries are $a_{ij}=1$ if a directed link from vertex $i$ to vertex $j$ is there, and $a_{ij}=0$ otherwise. The local constraints $\{C_a\}$ are now the two sets of out-degrees and in-degrees of all vertices defined in Eqs.(\ref{eq_controlkout}) and (\ref{eq_controlkin}), i.e. the \emph{out-degree sequence} $\{k^{out}_i\}$ and the \emph{in-degree sequence} $\{k^{in}_i\}$. 
In Appendix \ref{app_bdn} we show how the randomization method enables in this case to obtain the expectation value $\langle X\rangle$ of a property $X$ across the maximally random ensemble of binary directed graphs with in-degree and out-degree sequences equal to the observed ones. When inspecting the properties of the ITN and its randomized variants,  the useful independent variables are now the values $\{k^{out}_i\}$ and $\{k^{in}_i\}$ (or combinations of them), since they are the special quantities $X$ whose expected value $\langle X\rangle$ coincides with the observed one by construction. Again, we first consider the 2002 snapshot of the completely aggregated ITN (Sections \ref{sec_bdn_annd} and \ref{sec_bdn_clustering}), then track the temporal evolution of the results backwards (Section \ref{sec_bdn_evolution}), and finally perform a disaggregated analysis in Section \ref{sec_bdn_dis}.

\subsection{Directed average nearest neighbor degrees\label{sec_bdn_annd}}
We start with the analysis of the binary directed trade network aggregated over all commodities ($c=0$).
Therefore, in the following formulas, we set $\mathbf{A}\equiv\mathbf{A}^0$.
The average nearest neighbor degree of a vertex in a directed graph can be generalized in four ways from its undirected analogue. We thus obtain the quantities
\begin{eqnarray}
k_{i}^{in/in}&\equiv&\frac{\sum_{j\ne i}a_{ji}k_{j}^{in}}{k_{i}^{in}}
=\frac{\sum_{j\ne i}\sum_{k\ne j}a_{ji}a_{kj}}{\sum_{j\ne i}a_{ji}}\label{eq_bdn_kinin}\\
k_{i}^{in/out}&\equiv&\frac{\sum_{j\ne i}a_{ji}k_{j}^{out}}{k_{i}^{in}}
=\frac{\sum_{j\ne i}\sum_{k\ne j}a_{ji}a_{jk}}{\sum_{j\ne i}a_{ji}}\label{eq_bdn_kinout}
\\
k_{i}^{out/in}&\equiv&\frac{\sum_{j\ne i}a_{ij}k_{j}^{in}}{k_{i}^{out}}
=\frac{\sum_{j\ne i}\sum_{k\ne j}a_{ij}a_{kj}}{\sum_{j\ne i}a_{ij}}\label{eq_bdn_koutin}\\
k_{i}^{out/out}&\equiv&\frac{\sum_{j\ne i}a_{ij}k_{j}^{out}}{k_{i}^{out}}
=\frac{\sum_{j\ne i}\sum_{k\ne j}a_{ij}a_{jk}}{\sum_{j\ne i}a_{ij}}\label{eq_bdn_koutout}
\end{eqnarray}
In the above expressions, indirect interactions due to the concatenation of pairs of edges are taken into account according to their directionality, as clear from the presence of products of the type $a_{ij}a_{kl}$.
A fifth possibility is an aggregated measure based on the total degree $k^{tot}_i\equiv k^{in}_i+k^{out}_i$ of vertices:
\begin{equation}
k_{i}^{tot/tot}\equiv\frac{\sum_{j\ne i}(a_{ij}+a_{ji})k_{j}^{tot}}{k_{i}^{tot}}
\end{equation}
The latter is a useful one to start with, as it provides a simpler analogue to the undirected quantity $k^{nn}_i$ we have already studied. 
At the same time, it must be noted that the two quantities are not trivially related since the total directed properties $k^{tot}_i$ and $k_{i}^{tot/tot}$ carry more information than the corresponding undirected ones $k_i$ and $k^{nn}_i$, the difference being the local \emph{reciprocity structure} of the network \cite{myreciprocity}. 
To see this, note that $k^{tot}_i=k_i+k_{i}^{\leftrightarrow}$, where $k_{i}^{\leftrightarrow}\equiv\sum_{j\ne i}a_{ij}a_{ji}$ (if $a_{ij}$ is the adjacency matrix of the directed network) is the \emph{reciprocated degree} of vertex $i$, defined as the number of bidirectional links reaching $i$ \cite{myreciprocity,ZL,ZS}. This quantity represents the number of trade partners, acting simultaneously as importers and exporters, of country $i$. Therefore, studying total directed quantities also allows to assess whether the reciprocity structure of the directed network changes the results obtained in the undirected case (similar considerations apply to the directed clustering coefficients we introduce below).

\begin{figure}[t]
\includegraphics[width=.45\textwidth]{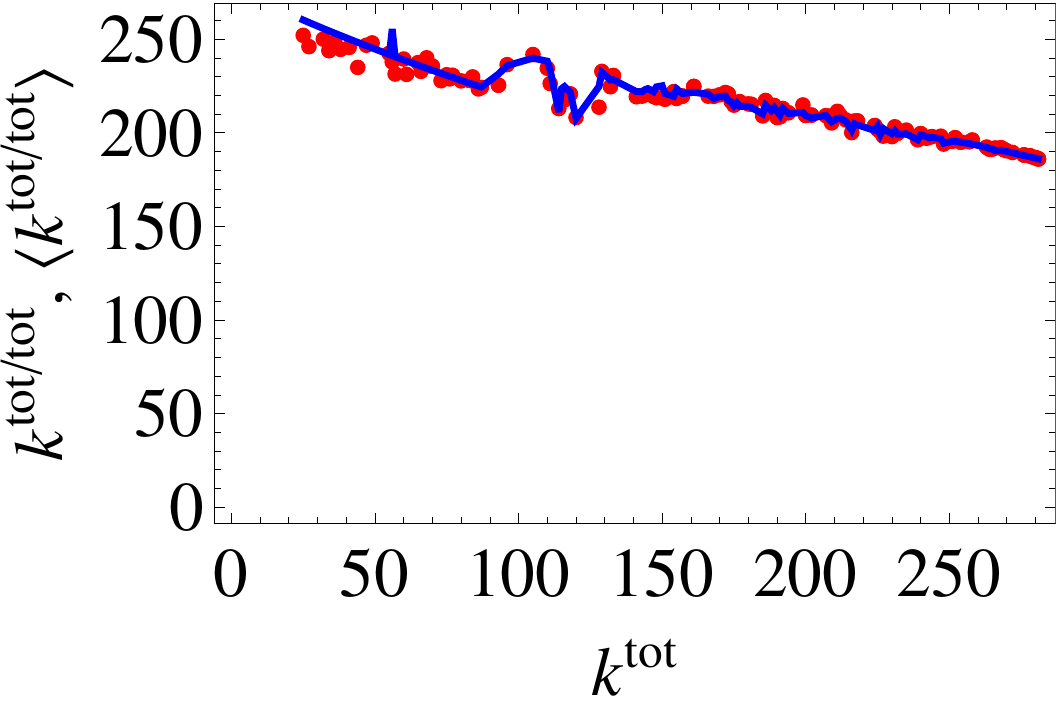}
\caption{\label{fig_bdn_knn} (Color online) Total average nearest neighbor degree $k^{tot/tot}_i$ versus total degree $k^{tot}_i$ in the 2002 snapshot of the real binary directed ITN (red points), and corresponding average over the maximum-entropy ensemble with specified out-degrees and in-degrees (blue solid curve).}
\end{figure}

In Fig.~\ref{fig_bdn_knn} we plot $k_{i}^{tot/tot}$ as a function of $k_{i}^{tot}$ for the 2002 snapshot of the binary directed ITN. The trend shown does not differ substantially from its undirected counterpart we observed in Fig.~\ref{fig_bun_knn}. In particular, we obtain a similar disassortative character of the correlation profile. Importantly, we find again a good agreement between the empirical quantity and its expected value $\langle k_{i}^{tot/tot}\rangle$ under the null model (obtained as in Appendix \ref{app_bdn}).
In Fig.~\ref{fig_bdn_knndir} we show a more refined analysis by considering all the four directed versions of the ANND defined in Eqs.(\ref{eq_bdn_kinin})-(\ref{eq_bdn_koutout}), as well as their expected values under the null model (see Appendix \ref{app_bdn}). We immediately see that all quantities still display a disassortative trend, with some differences in the ranges of observed values. Again, all the four empirical behaviors are in striking accordance with the null model, as the randomized curves (obtained as in Appendix \ref{app_bdn}) show. This means that both the decreasing trends
and the ranges of values displayed by all quantities are well reproduced by a collection of random graphs with the same average in-degrees and out-degrees as the real network.

\subsection{Directed clustering coefficients\label{sec_bdn_clustering}}
We now consider the directed counterparts of the clustering coefficient defined in Eq.~(\ref{eq_bun_ck}). Again, there are four possible generalizations depending on whether the directed triangles involved are of the \emph{inward}, \emph{outward}, \emph{cyclic} or \emph{middleman} type \cite{Fagiolo2007pre}:
\begin{eqnarray}
c_{i}^{in}&\equiv&\frac{\sum_{j\ne i}\sum_{k\ne i,j}a_{ki}a_{ji}a_{jk}}
{k_{i}^{in}(k_{i}^{in}-1)}\label{eq_bdn_cin}\\
c_{i}^{out}&\equiv&\frac{\sum_{j\ne i}\sum_{k\ne i,j}a_{ik}a_{jk}a_{ij}}
{k_{i}^{out}(k_{i}^{out}-1)}\label{eq_bdn_cout}\\
c_{i}^{cyc}&\equiv&\frac{\sum_{j\ne i}\sum_{k\ne i,j}a_{ij}a_{jk}a_{ki}}
{k_{i}^{in}k_{i}^{out}-k_{i}^{\leftrightarrow}}\label{eq_bdn_ccyc}\\
c_{i}^{mid}&\equiv&\frac{\sum_{j\ne i}\sum_{k\ne i,j}a_{ik}a_{ji}a_{jk}}
{k_{i}^{in}k_{i}^{out}-k_{i}^{\leftrightarrow}}\label{eq_bdn_cmid}
\end{eqnarray}
The directed clustering coefficients are determined by indirect interactions of length $3$ according to their directionality, appearing as products of the type $a_{ij}a_{kl}a_{mn}$ in the above formulas. At the same time, since they always focus on three vertices only, they capture the local occurrence of particular \emph{network motifs} \cite{motifs} of order $3$. 
A fifth aggregated measure, based on all possible directions, is
\begin{equation}
c_{i}^{tot}\equiv\frac{\sum_{j\ne i}\sum_{k\ne i,j}(a_{ij}+a_{ji})(a_{jk}+a_{kj})(a_{ki}+a_{ik})}
{2\big[k_{i}^{tot}(k_{i}^{tot}-1)-2 k_{i}^{\leftrightarrow}\big]}
\label{eq_bdn_ctot}
\end{equation}
As for $k^{tot/tot}_i$, the latter definition is a good starting point for a comparison with the undirected case. In Fig.~\ref{fig_bdn_ck} we show $c^{tot}_i$ and $\langle c^{tot}_i\rangle$ (see Appendix \ref{app_bdn}) as a function $k^{tot}_i$ for our usual snapshot. We see no fundamental difference with respect to Fig.~\ref{fig_bun_ck}. Again, the randomized quantity does not deviate significantly from the empirical one.

\begin{figure}[b]
\includegraphics[width=.45\textwidth]{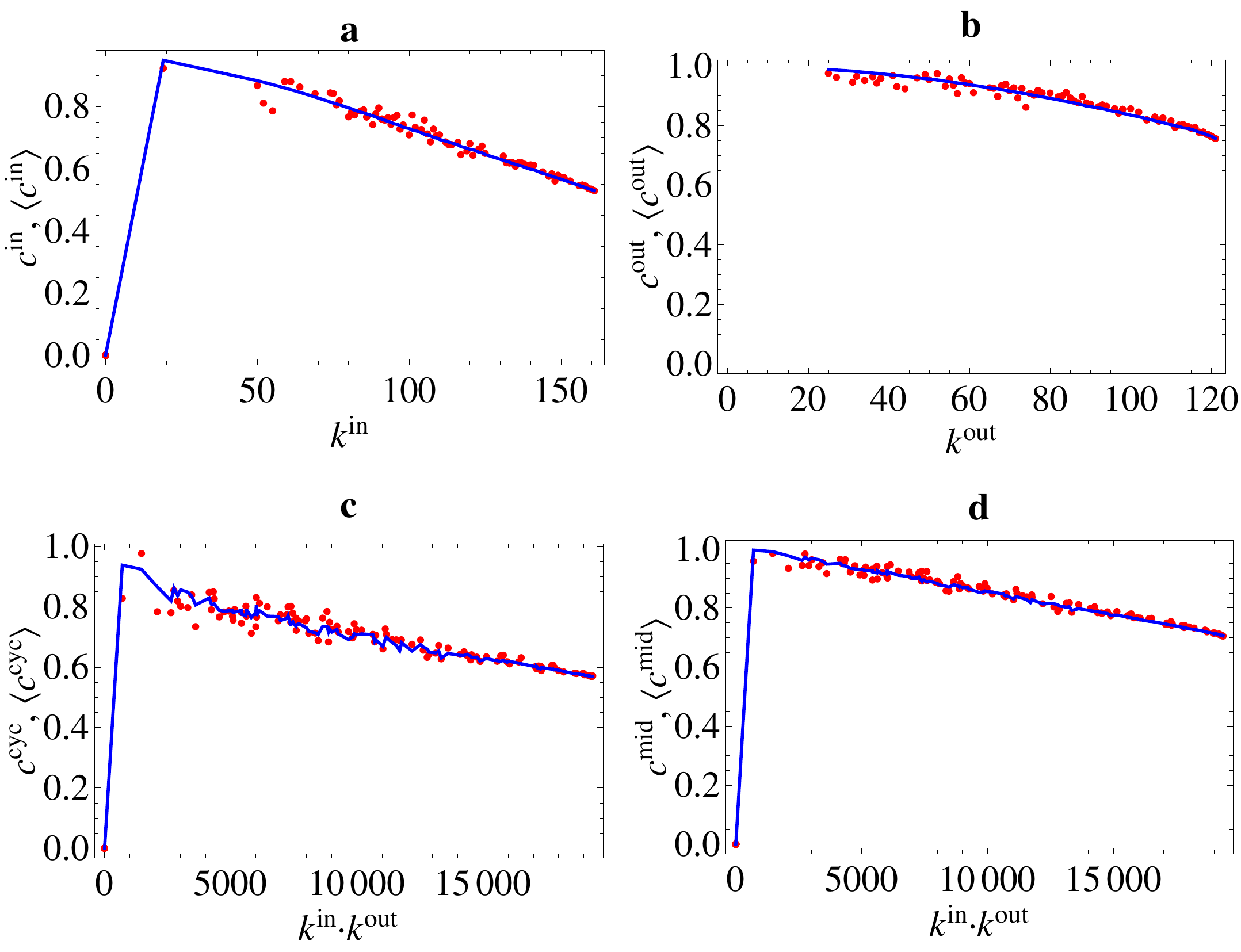}
\caption{\label{fig_bdn_cdir} (Color online) Directed clustering coefficients versus vertex degrees in the 2002 snapshot of the real binary directed ITN (red points), and corresponding averages over the maximum-entropy ensemble with specified out-degrees and in-degrees (blue solid curves). 
\textbf{a)} $c^{in}_i$ versus $k^{in}_i$. 
\textbf{b)} $c^{out}_i$ versus $k^{out}_i$. \textbf{c)} $c^{cyc}_i$ versus $k^{in}_i\cdot k^{out}_i$. 
\textbf{d)} $c^{mid}_i$ versus $k^{in}_i\cdot k^{out}_i$.}
\end{figure}

We now turn to the four directed clustering coefficients defined in Eqs.(\ref{eq_bdn_cin})-(\ref{eq_bdn_cmid}). We show these quantities in Fig.~\ref{fig_bdn_cdir} as functions of different combinations of $k^{in}_i$ and $k^{out}_i$, depending on the particular definition. As for the directed ANND, we observe some variability in the range of observed clustering values. However, all the quantities are again in accordance with the expected ones under the null model (see Appendix \ref{app_bdn}).
\begin{figure}[b]
\includegraphics[width=.45\textwidth]{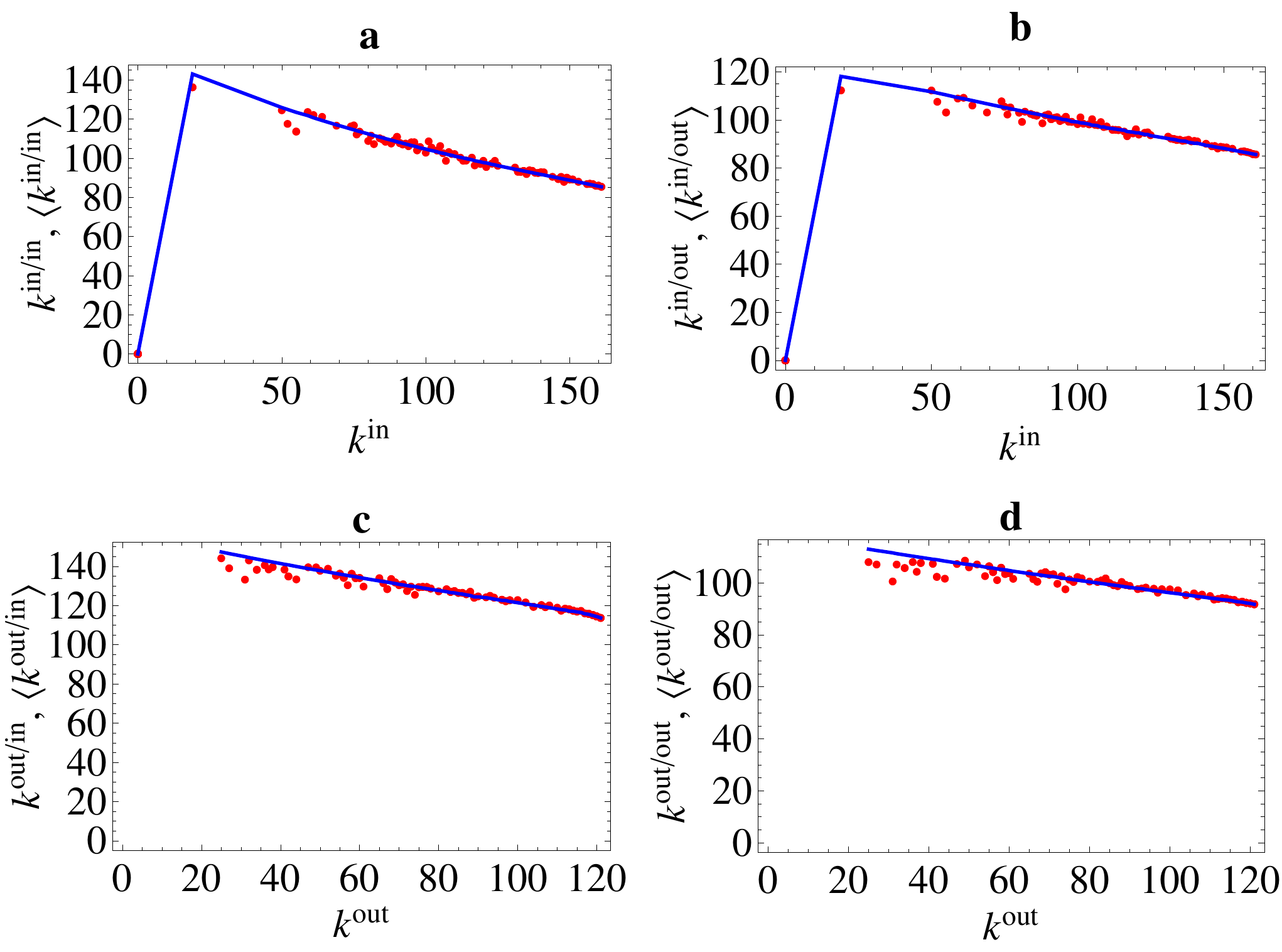}
\caption{\label{fig_bdn_knndir} (Color online) Directed average nearest neighbor degrees versus vertex degrees in the 2002 snapshot of the real binary directed ITN (red points), and corresponding averages over the maximum-entropy ensemble with specified out-degrees and in-degrees (blue solid curves).
\textbf{a)} $k^{in/in}_i$ versus $k^{in}_i$.
\textbf{b)} $k^{in/out}_i$ versus $k^{in}_i$.
\textbf{c)} $k^{out/in}_i$ versus $k^{out}_i$.
\textbf{d)} $k^{out/out}_i$ versus $k^{out}_i$.}
\end{figure}
\begin{figure}[t]
\includegraphics[width=.45\textwidth]{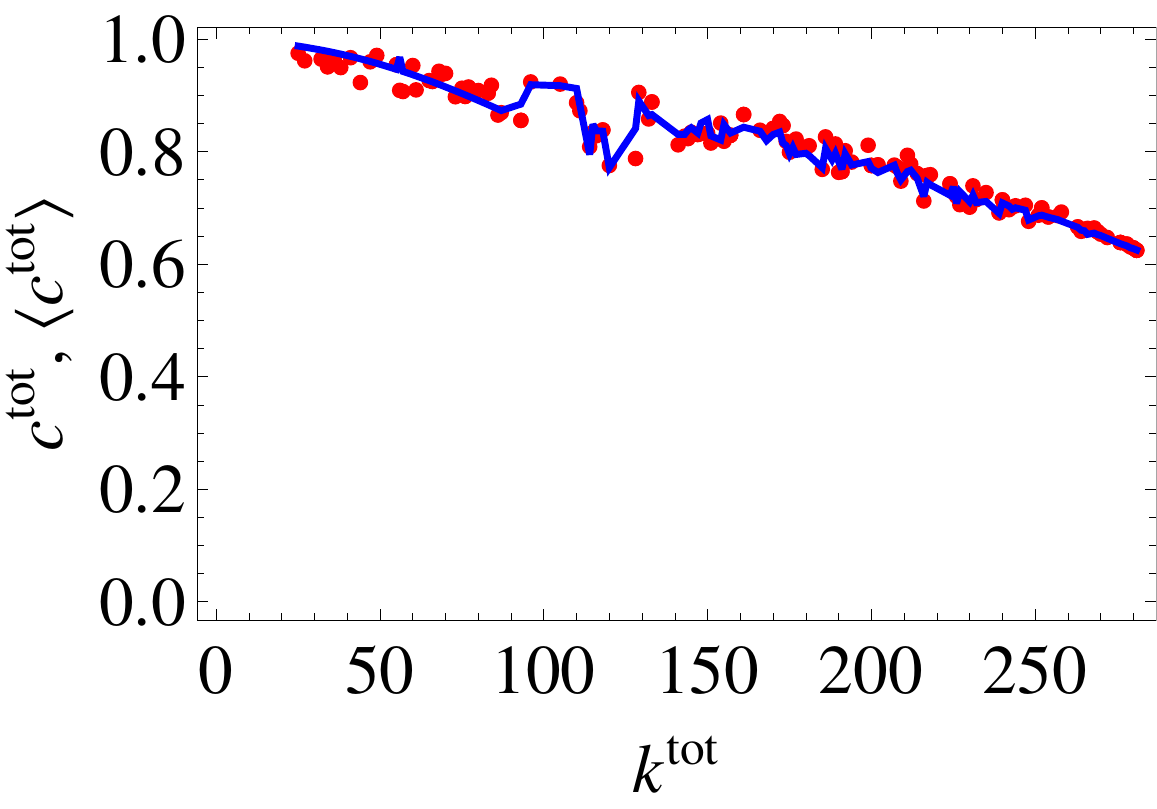}
\caption{\label{fig_bdn_ck} (Color online) Total clustering coefficient $c^{tot}_i$ versus total degree $k^{tot}_i$ in the 2002 snapshot of the real binary directed ITN (red points), and corresponding average over the maximum-entropy ensemble with specified out-degrees and in-degrees (blue solid curve).}
\end{figure}

\subsection{Evolution of binary directed properties\label{sec_bdn_evolution}}
We now track the temporal evolution of the above results by performing, for each year in our time window, an analysis similar to that reported in sec.\ref{sec_bun_evolution} for the undirected case.

\begin{figure}[t]
\includegraphics[width=.45\textwidth]{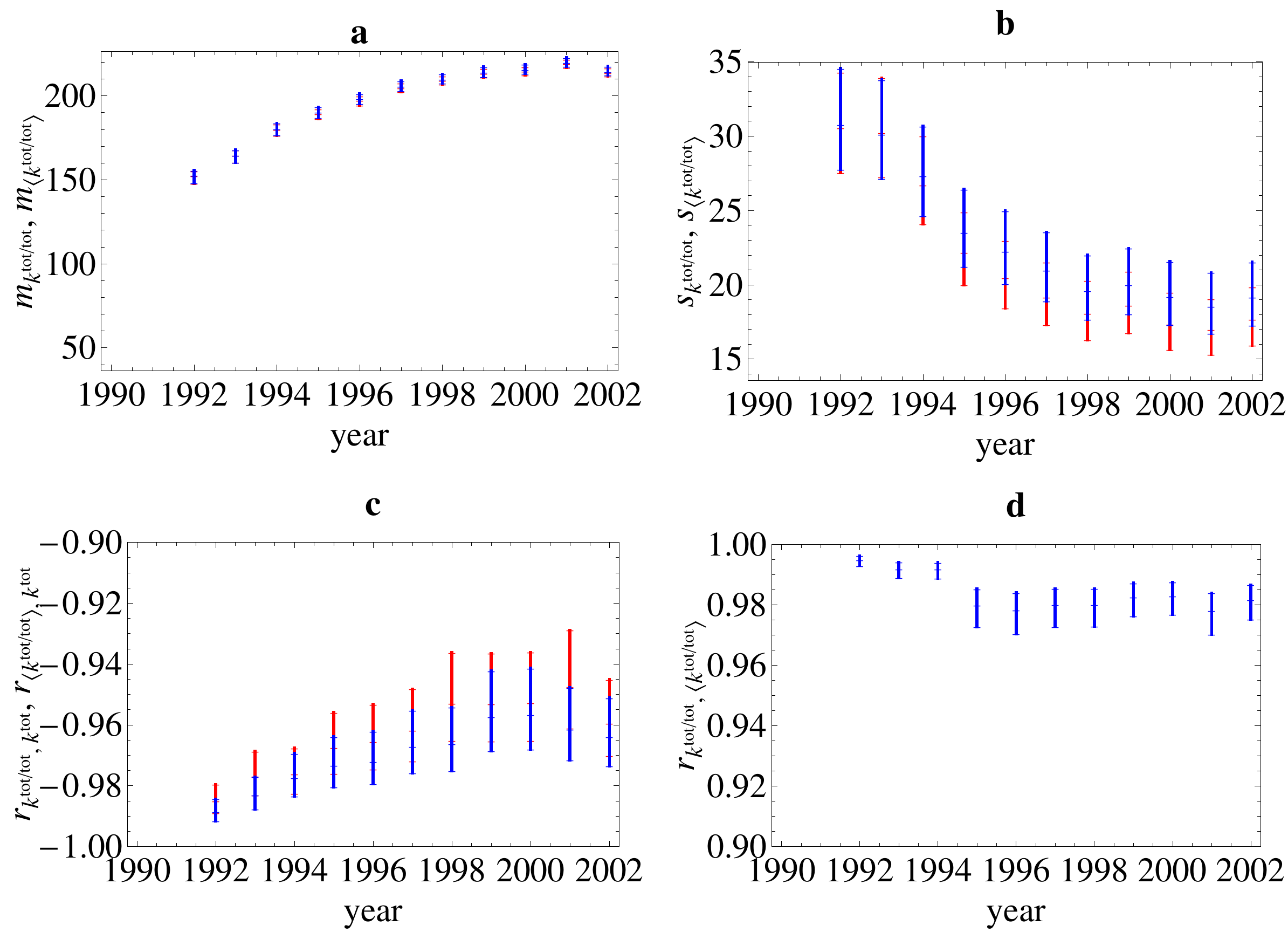}
\caption{\label{fig_bdn_ktot_t}
(Color online) Temporal evolution of the properties of the total average nearest neighbor degree $k^{tot/tot}_i$ in the 1992-2002 snapshots of the real binary directed ITN and of the corresponding null model with specified out-degrees and in-degrees. \textbf{a)} average of $k^{tot/tot}_i$ across all vertices (red: real data; blue: null model, indistinguishable from real data). \textbf{b)} standard deviation of $k^{tot/tot}_i$ across all vertices (red: real data; blue: null model, overlapping with real data). \textbf{c)} correlation coefficient between $k^{tot/tot}_i$ and $k^{tot}_i$ (red: real data; blue: null model, overlapping with real data). \textbf{d)} correlation coefficient between $k^{tot/tot}_i$ and $\langle k^{tot/tot}_i\rangle$. The $95\%$ confidence intervals of all quantities are represented as vertical bars.}
\end{figure}

We start by showing the evolution of the total average nearest neighbor degree $k^{tot/tot}_i$ in the four panels of Fig.~\ref{fig_bdn_ktot_t}, where we plot the same properties considered previously for the undirected ANND in Fig.~\ref{fig_bun_knn_t}.
We find that the temporal evolution of the average (Fig.~\ref{fig_bdn_ktot_t}a) and standard deviation (Fig.~\ref{fig_bdn_ktot_t}b) of $k^{tot/tot}_i$ is essentially the same as that of the undirected $k^{nn}_i$, apart from  differences in the range of values.
Similarly, the correlation coefficients between $k^{tot/tot}_i$ and $k^{tot}_i$ (Fig.~\ref{fig_bdn_ktot_t}c), $\langle k^{tot/tot}_i\rangle$ and $\langle k^{tot}_i\rangle=k^{tot}_i$ (Fig.~\ref{fig_bdn_ktot_t}c), $k^{tot/tot}_i$ and $\langle k^{tot/tot}_i\rangle$ (Fig.~\ref{fig_bdn_ktot_t}d) mimic their undirected counterparts, confirming that the perfect accordance between $k^{tot/tot}_i$ and $\langle k^{tot/tot}_i\rangle$ is stable over time, and that the disassortative trend of $k^{tot/tot}_i$ as a function of $k^{tot}_i$ (Fig.~\ref{fig_bdn_knn}) is always completely explained by the null model.

\begin{figure}[b]
\includegraphics[width=.44\textwidth]{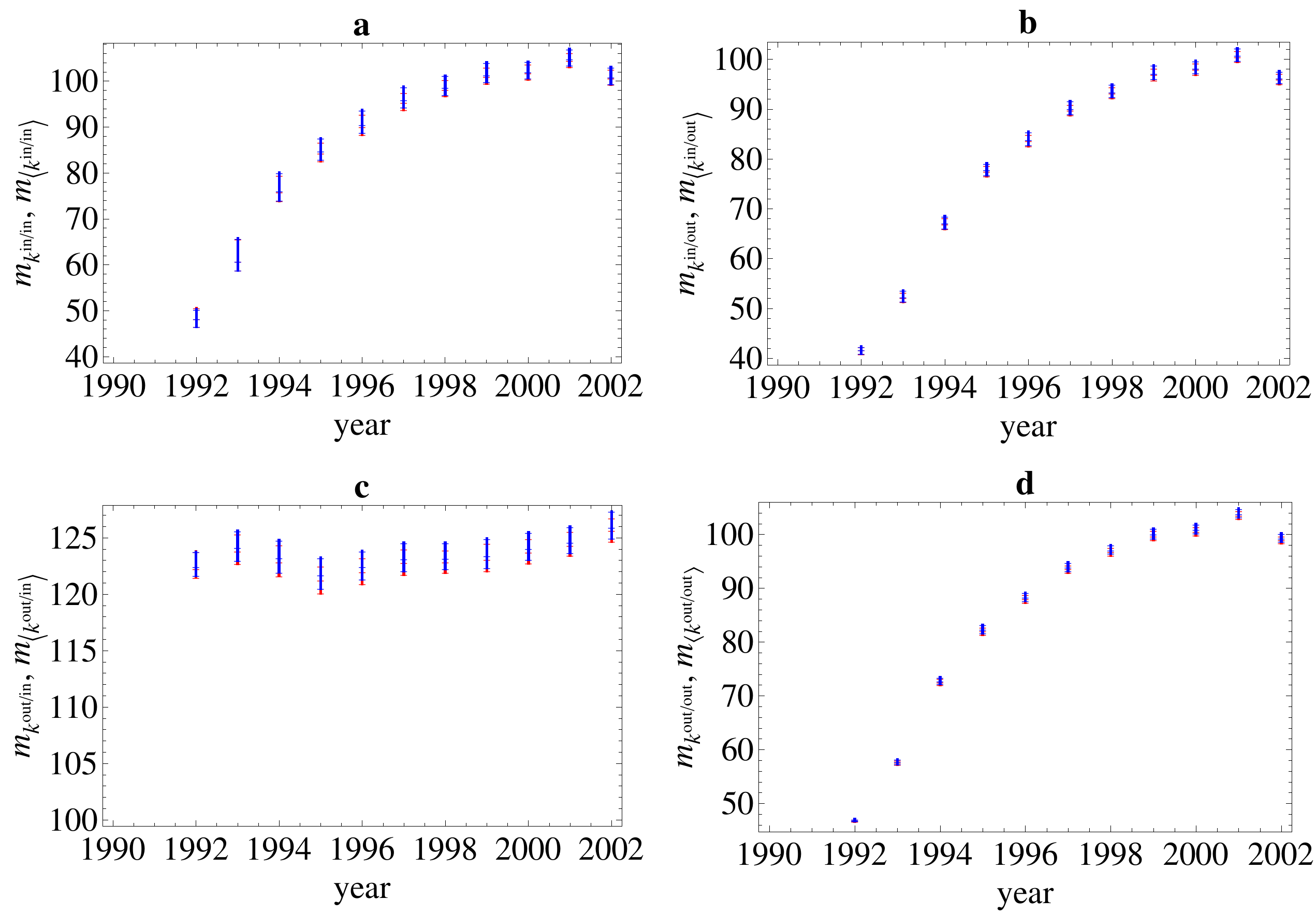}
\caption{\label{fig_bdn_knndir_t} (Color online) Averages and their $95\%$ confidence intervals (across all vertices) of the directed average nearest neighbor degrees in the 1992-2002 snapshots of the real binary directed ITN (red), and corresponding averages over the null model with specified out-degrees and in-degrees (blue, indistinguishable from real data).
\textbf{a)} average of $k^{in/in}_i$; 
\textbf{b)} average of $k^{in/out}_i$;
\textbf{c)} average of $k^{out/in}_i$;
\textbf{d)} average of $k^{out/out}_i$.}
\end{figure}

\begin{figure}[t]
\includegraphics[width=.45\textwidth]{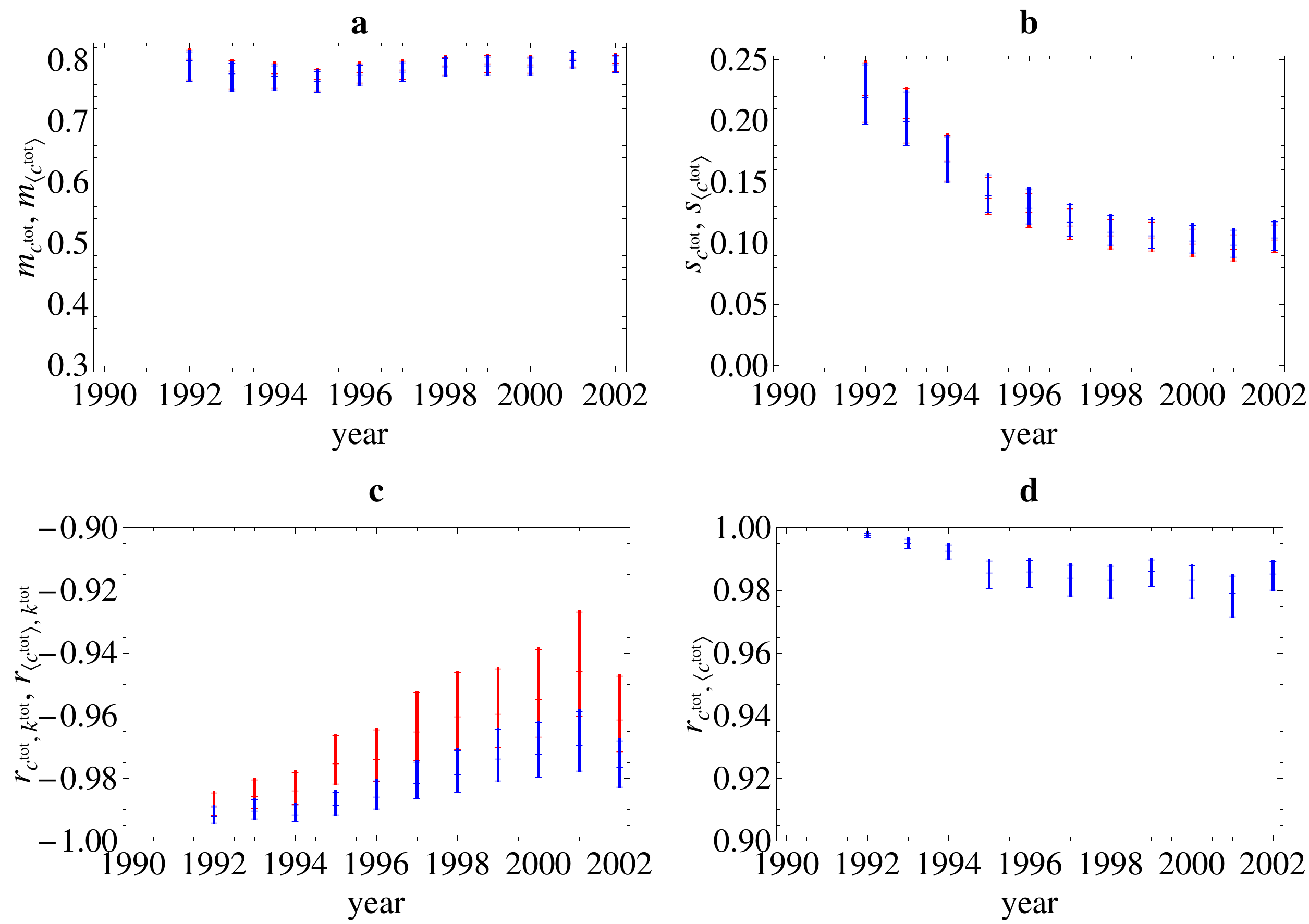}
\caption{\label{fig_bdn_ctot_t}
(Color online) Temporal evolution of the properties of the total clustering coefficient $c^{tot}_i$ in the 1992-2002 snapshots of the real binary directed ITN and of the corresponding null model with specified out-degrees and in-degrees. 
\textbf{a)} average of $c^{tot}_i$ across all vertices (red: real data; blue: null model, indistinguishable from real data). 
\textbf{b)} standard deviation of $c^{tot}_i$ across all vertices (red: real data; blue: null model, indistinguishable from real data). 
\textbf{c)} correlation coefficient between $c^{tot}_i$ and $k^{tot}_i$ (red, upper symbols: real data; blue, lower symbols: null model). 
\textbf{d)}  correlation coefficient between $c^{tot}_i$ and $\langle c^{tot}_i\rangle$. The $95\%$ confidence intervals of all quantities are represented as vertical bars.}
\end{figure}
\begin{figure}[b]
\includegraphics[width=.45\textwidth]{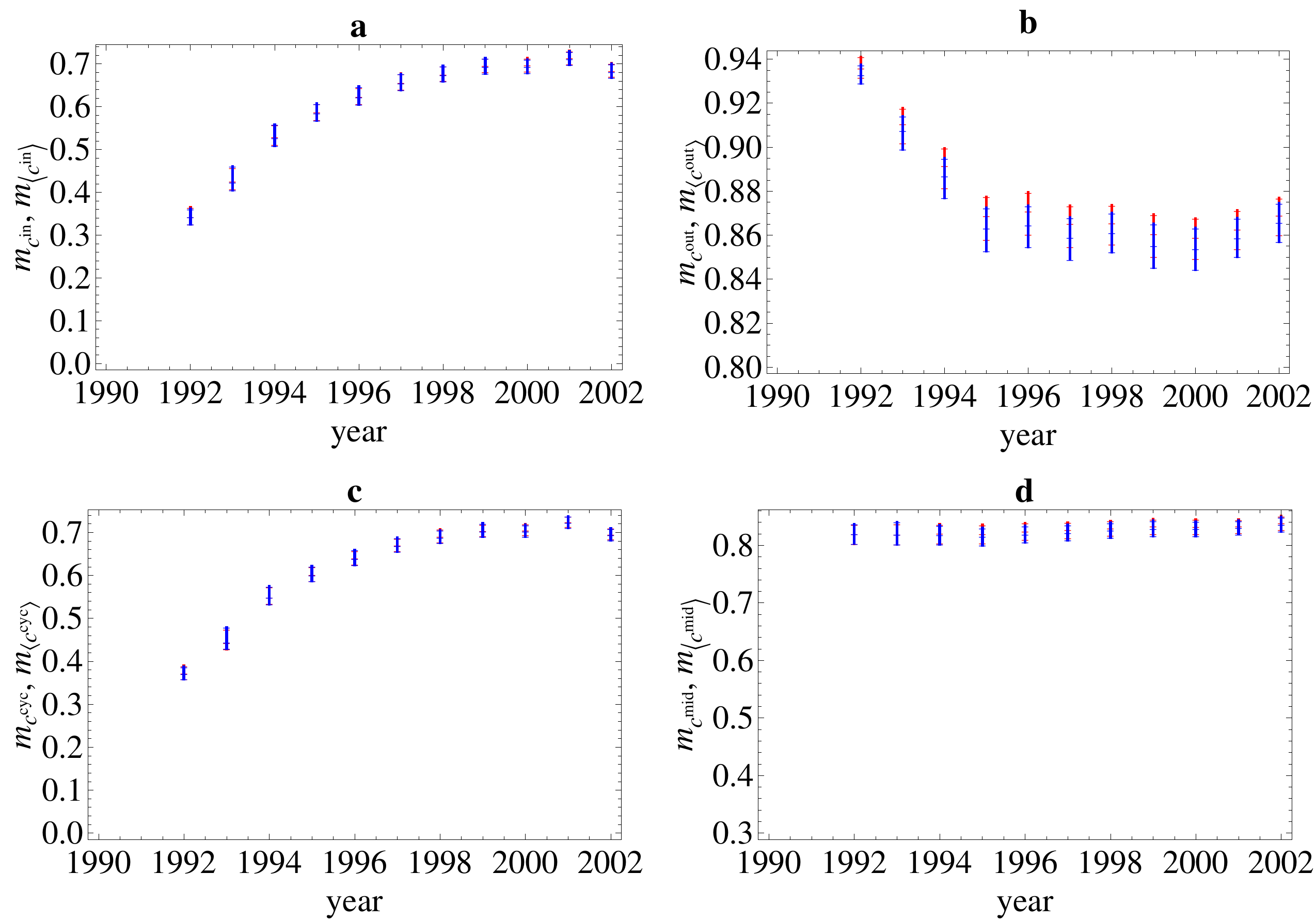}
\caption{\label{fig_bdn_cdir_t} (Color online) Averages and their $95\%$ confidence intervals (across all vertices) of the directed clustering coefficients  in the 1992-2002 snapshots of the real binary directed ITN (red), and corresponding averages over the null model with specified out-degrees and in-degrees (blue, indistinguishable from real data).
\textbf{a)} $c^{in}_i$;
\textbf{b)}  $c^{out}_i$;
\textbf{c)}  $c^{cyc}_i$;
\textbf{d)}  $c^{mid}_i$.}
\end{figure}

We now consider the four directed variants $k^{in/in}_i$, $k^{in/out}_i$, $k^{out/in}_i$, $k^{out/out}_i$. For brevity, for these quantities we only show the evolution of the average values, which are reported in Fig.~\ref{fig_bdn_knndir_t}. We find that the overall behavior previously reported for the average of $k^{tot/tot}_i$ (Fig.~\ref{fig_bdn_ktot_t}a) is not reflected in the individual trends of the four directed versions of the ANND. In particular, the averages of $k^{in/in}_i$ (Fig.~\ref{fig_bdn_knndir_t}a), $k^{in/out}_i$ (Fig.~\ref{fig_bdn_knndir_t}b) and $k^{out/out}_i$ (Fig.~\ref{fig_bdn_knndir_t}d) increase over a downward-shifted but wider range of values than that of  $k^{tot/tot}_i$, whereas the average of $k^{out/in}_i$ (Fig.~\ref{fig_bdn_knndir_t}c) is almost constant in time. The moderately increasing average of $k^{tot/tot}_i$ is therefore the overall result of a combination of different trends followed by the underlying directed quantities, some of these trends being strongly increasing and some being almost constant.
Therefore we find the important result that \emph{there is a substantial loss of information in passing from the inherently directed quantities to the undirected or symmetrized ones}.
Still, when we compare the empirical trends of the directed quantities with the randomized ones, we find an almost perfect agreement. This implies that even the finer structure of directed correlation profiles, as well as their evolution, is reproduced in great detail by controlling for the local directed topological properties alone.

\begin{figure}[t]
\includegraphics[width=.45\textwidth]{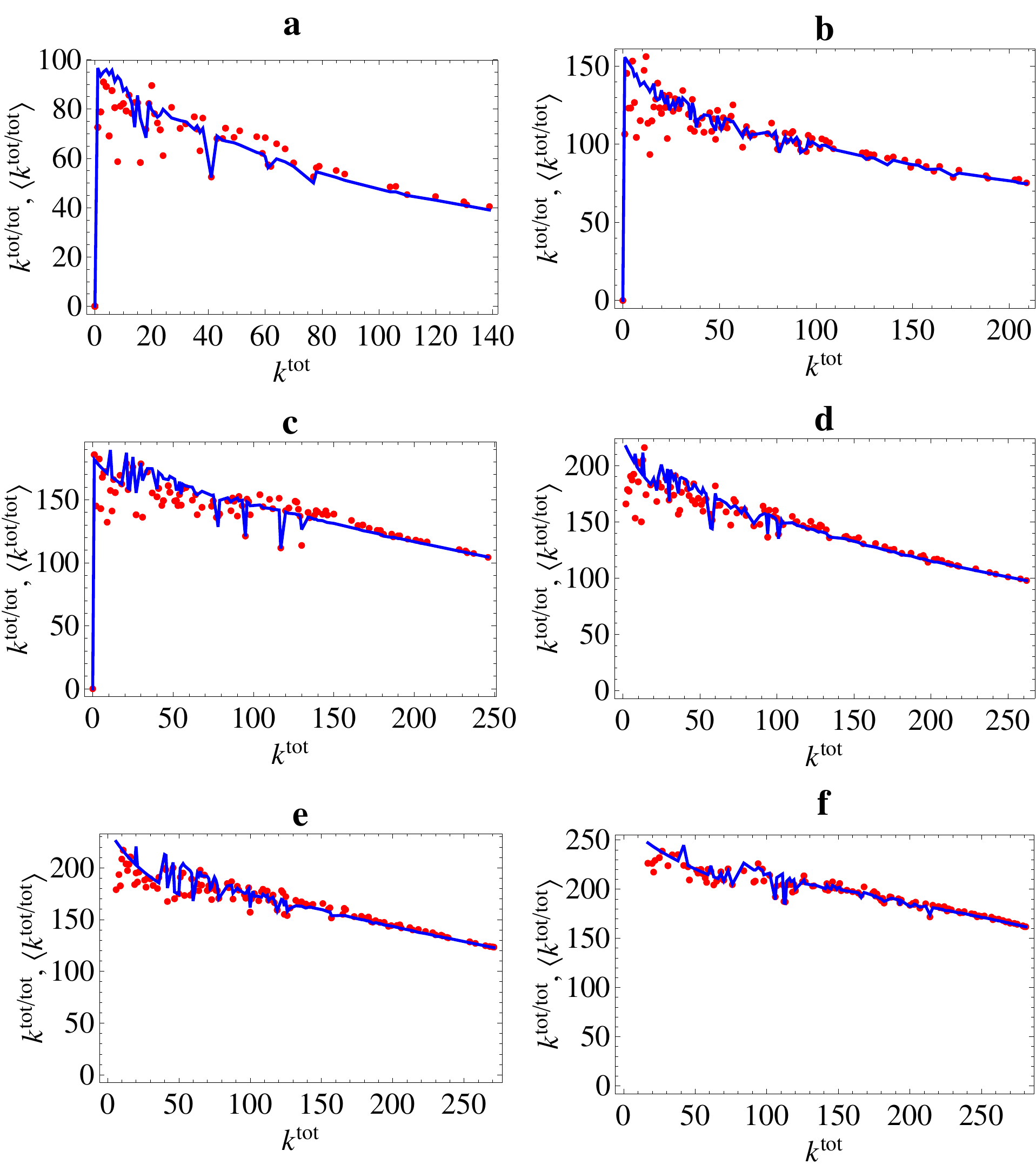}
\caption{\label{fig_bdn_dis1}
(Color online) Total average nearest neighbor degree $k^{tot/tot}_i$ versus total degree $k^{tot}_i$ in the 2002 snapshots of the commodity-specific (disaggregated) versions of the real binary directed ITN (red points), and corresponding average over the maximum-entropy ensemble with specified out-degrees and in-degrees (blue solid curves).
\textbf{a)} commodity 93; 
\textbf{b)} commodity 09; 
\textbf{c)} commodity 39; 
\textbf{d)} commodity 90; 
\textbf{e)} commodity 84; 
\textbf{f)} aggregation of the top 14 commodities (see Table \ref{table} for details). From a) to f), the intensity of trade and level of aggregation increases.} 
\end{figure}

The same analysis is shown for the total clustering coefficient $c^{tot}_i$ in Fig.~\ref{fig_bdn_ctot_t}, and for the four directed variants $c^{in}_i$, $c^{out}_i$, $c^{cyc}_i$, $c^{mid}_i$ in Fig.~\ref{fig_bdn_cdir_t}. Again, we find that the four temporal trends involving the overall quantity $c^{tot}_i$ (Fig.~\ref{fig_bdn_ctot_t}) replicate what we have found for its undirected counterpart $c_i$ (shown previously in Fig.~\ref{fig_bun_c_t}). When we consider the four inherently directed quantities (Fig.~\ref{fig_bdn_cdir_t}), we find that the averages of $c^{in}_i$ (Fig.~\ref{fig_bdn_cdir_t}a) and $c^{cyc}_i$ (Fig.~\ref{fig_bdn_cdir_t}c) display an increasing trend, whereas the average of $c^{mid}_i$ (Fig.~\ref{fig_bdn_cdir_t}d) is constant and that of $c^{out}_i$ (Fig.~\ref{fig_bdn_cdir_t}b) is even decreasing. When aggregated, these different trends give rise to the constant behavior of the average $c^{tot}_i$, \emph{which is therefore not representative of the four underlying directed quantities.} This also means that, similarly to what we found for the ANND, there is a substantial loss of information in passing from the directed to the undirected description of the binary ITN. However, even the fine-level differences among the directed clustering patterns are still completely reproduced by the null model.

\begin{figure}[t]
\includegraphics[width=.45\textwidth]{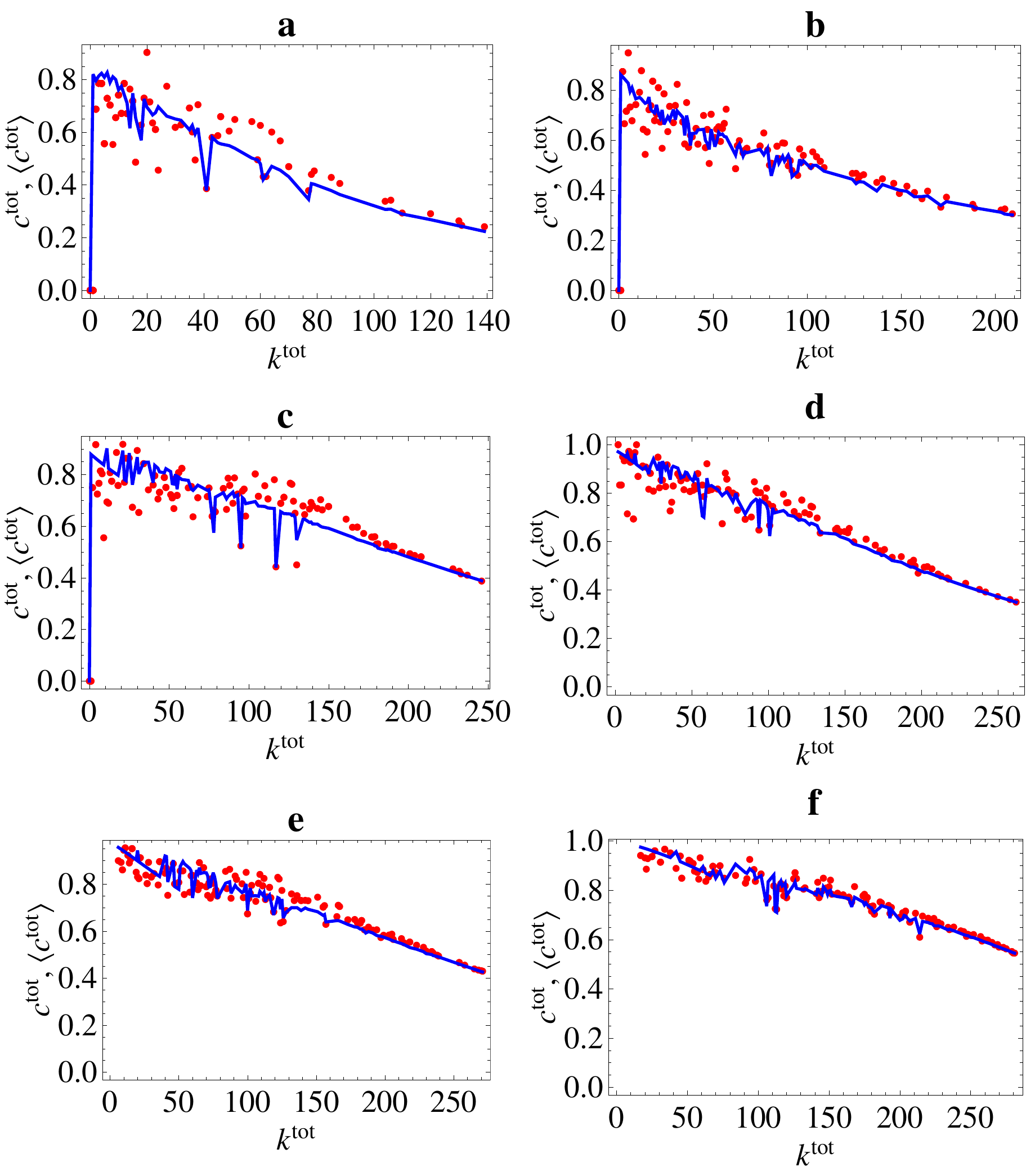}
\caption{\label{fig_bdn_dis2}
(Color online) Total clustering coefficient $c^{tot}_i$ versus total degree $k^{tot}_i$ in the 2002 snapshots of the commodity-specific (disaggregated) versions of the real binary directed ITN (red points), and corresponding average over the maximum-entropy ensemble with specified our-degrees and in-degrees (blue solid curves).
\textbf{a)} commodity 93; 
\textbf{b)} commodity 09; 
\textbf{c)} commodity 39; 
\textbf{d)} commodity 90; 
\textbf{e)} commodity 84; 
\textbf{f)} aggregation of the top 14 commodities (see Table \ref{table} for details).
From a) to f), the intensity of trade and level of aggregation increases.} 
\end{figure}

\subsection{Commodity-specific binary directed networks\label{sec_bdn_dis}}
We now study the binary directed ITN when disaggregated (commodity-specific) representations are considered. 
We repeat the analysis described above by setting $\mathbf{A}\equiv\mathbf{A}^c$ with $c>0$.
For brevity, we report our analysis of the 6 commodities described in Section \ref{sec_bun_dis} and selected from the top 14 categories listed in table \ref{table} (again, we found similar results for all commodities). Together with the aggregated binary directed ITN already described, these commodity classes form a set of 7 different cases ordered by increasing trade intensity and level of commodity aggregation.

In Figs.~\ref{fig_bdn_dis1} and  \ref{fig_bdn_dis2} we show the behavior of the total average nearest neighbor degree and total clustering coefficient for the 2002 snapshots of the 6 selected commodity-specific networks. 
When compared with Figs.~\ref{fig_bdn_knn} and \ref{fig_bdn_ck}, the plots confirm what we have found in Section \ref{sec_bun_dis} for the binary undirected case. In particular, the behavior displayed by the ANND and clustering in the commodity-specific networks becomes less and less noisy as more intensely traded commodities, and higher levels of aggregation, are considered. Accordingly, the agreement between real and randomized networks increases, but the accordance is already remarkable in commodity-specific  networks, even the sparsest and least aggregated ones.
These results confirm that, irrespective of the level of commodity resolution and trade volume, the directed degree sequences completely characterize the topology of the binary directed representations of the ITN.

\section{Conclusions}
All the above results clearly imply that, in the undirected as well as the directed case, for all the years considered, and across different commodity classes, the disassortativity and clustering profiles observed in the real binary ITN arise as natural outcomes rather than genuine correlations, once the local topological properties are fixed to their observed values.
Therefore we can conclude that the higher-order patterns observed in all the binary representations of the ITN, as well as their temporal evolution, are completely explained by local constraints alone.
This means that the (undirected/directed) degree sequence of the ITN is maximally informative, since its knowledge systematically conveys a full picture of the binary topology of the network. 
These results have important consequences for economic  models of trade. In particular, they suggest that the ITN topology should become one of the main focuses of international-trade theories. While most of the literature concerned with modeling international trade has focused on the problem of reproducing the magnitude of nonzero trade volumes (the most important example being gravity models \cite{Fagiolo2010jeic}), much less emphasis has been put on correctly replicating the binary topology of the ITN, i.e. understanding the determinants of the process governing the creation of a link. However, our results clearly show that the purely topological structural properties (and in particular the degree sequence) of the ITN carry a significant amount of information.

A first step in reproducing the ITN topology is the model in Ref.~\cite{Garla2004}, where the probability $p_{ij}$ of a trade relationship between two countries $i$ and $j$ was modeled as a function of the GDP values of the countries themselves, and all the topological properties of the network were successfully replicated. 
Interestingly, the form of that function coincides with the connection probability of the null model considered here, shown in Appendix \ref{app_bun} in Eq.~(\ref{eq_bun_pij}), where the role of the Lagrange multiplier $x_i$ associated with $k_i$  is played by the GDP of country $i$. Indeed, an approximately monotonic relationship between GDP and degree has been observed \cite{Garla2004}, providing a connection between these two results.
From another perspective, the above remark also means that the accordance between the real ITN as a binary undirected network and its randomized counterpart is replicated under an alternative null model, that controls for the empirical values of the GDP rather than for the degree sequence. 
The importance of reproducing the binary topology of trade is reinforced by the analysis of the ITN as a weighted network with local constraints, as we show in the following paper \cite{part2}.

\begin{acknowledgments}
D.G. acknowledges financial support from the European Commission 6th FP (Contract CIT3-CT-2005-513396), Project: DIME - Dynamics of Institutions and Markets in Europe. This work was also supported by the Dutch Econophysics Foundation (Stichting Econophysics, Leiden, Netherlands) with funds from Duyfken Trading Knowledge BV, Amsterdam, Netherlands.
\end{acknowledgments}

\appendix
\section{The randomization method\label{sec_method}}
Given a real network with $N$ vertices, there are various ways to generate a family of randomized variants of it \cite{Holland_Leinhardt_1976,Kannan_etal_1999,Katz_Powell_1957,Rao_etal_1996,Roberts_2000,Shen-Orr_etal_2002,Snijders_1991,maslov,msz,chunglu,myrandomization}. The most popular one is the \emph{local rewiring algorithm} proposed by Maslov and Sneppen \cite{maslov,msz}. In this method, one starts with the real network and generates a series of randomized graphs by iterating a fundamental rewiring step that preserves the desired properties. In the binary undirected case, where one wants to preserve the degree of every vertex, the steps are as follows: choose two edges, say $(i,j)$ and $(k,l)$; rewire these connections by swapping the end-point vertices and producing two new candidate edges, say $(i,l)$ and $(k,j)$; if these two new edges are not already present, accept them and delete the initial ones. After many iterations, this procedure generates a randomized variant of the original network, and by repeating this exercise a sufficiently large number of times, many randomized variants are obtained. By construction, all these variants have exactly the same degree sequence as the real-world network, but are otherwise random. In the directed and/or weighted case, extensions of the rewiring steps defined above can be introduced, even if with some caution \cite{serrc07,Opsahl_etal_2008}. Maslov and Sneppen's method allows one to check whether the enforced properties are partially responsible for the topological organization of the network. For instance, one can measure the degree correlations, or the clustering coefficient, across the randomized graphs and compare them with the empirical values measured on the real network. This method has been applied to various networks, including the Internet and protein networks \cite{maslov,msz}. Different webs have been found to be affected in very different ways by local constraints, making the problem interesting and not solvable \emph{a priori}.

The main drawback of the local rewiring algorithm is its computational requirements. Since the method is entirely computational, and analytical expressions for its results are not available, one needs to explicitly generate several randomized graphs, measure the properties of interest on each of them (and store their values), and finally perform an average. This average is an approximation for the actual expectation value over the entire set of allowed graphs. In order to have a good approximation, one needs to generate a large number $M$ of network variants. Thus, the time required to analyze the impact of local constraints on any structural property is $M$ times the time required to measure that property on the original network, plus the time required to perform many rewiring steps producing each of the $M$ randomized networks. The number of rewiring steps required to obtain a single randomized network is $O(L)$ where $L$ is the number of links \cite{maslov,msz,myrandomization}, and $O(L)=O(N)$ for sparse networks while $O(L)=O(N^2)$ for dense networks \footnote{It must be noted that the ITN is an unusually dense network. Density in the aggregate directed network indeed ranges from 0.23 in year 1992 to 0.56 in year 2001. Also product-specific networks are relatively dense. Density attains its maximum value for commodity 84 and its minimum for commodity 93. As a result, in the case of the ITN, applying the local rewiring algorithm computationally would have been extremely time consuming.}. Thus, if the time required to measure a given topological property on the original network is $O(N^\tau)$, the time required to measure the randomized value of the same property is $O(M\cdot L)+O(M\cdot N^\tau)$, which is $O(M\cdot N^\tau)$ as soon as $\tau \ge 2$.

A recently proposed alternative method, which is remarkably faster due to its analytical character, is based on the maximum-likelihood estimation of maximum-entropy models of graphs \cite{myrandomization}. In this method, one first specifies the desired set of local constraints $\{C_a\}$. Second, one writes down the analytical expression for the probability $P(\mathbf{G})$ that, subject to the constraints $\{C_a\}$, maximizes the entropy
\begin{equation}
S\equiv -\sum_\mathbf{G} P(\mathbf{G})\ln P(\mathbf{G})
\label{eq_entropy}
\end{equation}
where $\mathbf{G}$ denotes a particular graph in the ensemble, and $P(\mathbf{G})$ is the probability of occurrence of that graph. This probability defines the ensemble featuring the desired properties, and being maximally random otherwise. Depending on the particular description adopted, the graphs $\mathbf{G}$ can be either binary or weighted, and either directed or undirected. Accordingly, the sum in Eq.~(\ref{eq_entropy}), and in similar expressions shown later on, runs over all graphs of the type specified.
The formal solution to the entropy maximization problem can be written in terms of the so-called Hamiltonian $H(\mathbf{G})$, representing the energy (or cost) associated to a given graph $\mathbf{G}$. The Hamiltonian is defined as a linear combination of the specified constraints $\{C_a\}$:
\begin{equation}
H(\mathbf{G})\equiv\sum_a \theta_a C_a(\mathbf{G})
\label{eq_H}
\end{equation}
where $\{\theta_a\}$ are free parameters, acting as Lagrange multipliers controlling the expected values $\{\langle C_a\rangle\}$ of the constraints across the ensemble. The notation $C_a(\mathbf{G})$ denotes the particular value of the quantity $C_a$ when the latter is measured on the graph $\mathbf{G}$.
In terms of $H(\mathbf{G})$, the maximum-entropy graph probability  $P(\mathbf{G})$ can be shown to be
\begin{equation}
P(\mathbf{G})=\frac{e^{-H(\mathbf{G})}}{Z}
\end{equation}
where the normalizing quantity $Z$ is the \emph{partition function}, defined as
\begin{equation}
Z\equiv\sum_\mathbf{G} e^{-H(\mathbf{G})}
\label{eq_partition}
\end{equation}
Third, one maximizes the likelihood $P(\mathbf{G^*})$ to obtain the particular graph $\mathbf{G^*}$, which is the real-world network that one wants to randomize. This steps fixes the values of the Lagrange multipliers that finally allow to obtain the numerical values of the expected topological properties averaged over the randomized ensemble of graphs.
The particular values of the parameters $\{\theta_a\}$ that enforce the local constraints, as observed on the particular real network $\mathbf{G^*}$, are found by maximizing the log-likelihood
\begin{equation}
\lambda\equiv \ln P(\mathbf{G^*})=-H(\mathbf{G^*})-\ln Z
\label{eq_likelihood}
\end{equation}
to obtain the real network $\mathbf{G^*}$. It can be shown \cite{mylikelihood} that this is equivalent to the requirement that the ensemble average $\langle C_a\rangle$ of each constraint $C_a$ equals the empirical value measured on the real network:
\begin{equation}
\langle C_a\rangle=C_a(\mathbf{G^*})\quad \forall a
\end{equation}
Note that we generally adopted a simplified notation by writing $C_a^*$ (or even only $C_a$) instead of $C_a(\mathbf{G^*})$ for the empirically observed values of the constraints (see for instance Sections \ref{sec_data} and \ref{sec_controlling}).
Once the parameter values are found, they are inserted into the formal expressions yielding the expected value
\begin{equation}
\langle X\rangle\equiv \sum_\mathbf{G} X(\mathbf{G}) P(\mathbf{G})
\label{eq_X}
\end{equation}
of any (higher-order) property of interest $X$. The quantity $\langle X\rangle$ represents the average value of the property $X$ across the ensemble of random graphs with the same average (across the ensemble itself) constraints as the real network. 
In what follows we provide a detailed account of the expressions for the randomized properties appearing in our analysis.

Technically, while the local rewiring algorithm generates a \emph{microcanonical} ensemble of graphs, containing only those graphs for which the value of each constraint $C_a$ is exactly equal to the observed value $C_a(\mathbf{G}^*)$, the maximum-likelihood method generates an expanded \emph{grandcanonical} ensemble where all possible graphs with $N$ vertices are present, but where the ensemble average of each constraint $C_a$ is equal to the observed value $C_a(\mathbf{G}^*)$. One can show that the microcanonical approach converges to the grandcanonical one as the number of microcanonical randomization steps increases \cite{myrandomization}. However, the maximum-likelihood one is significantly faster.
Importantly, enforcing only local constraints implies that $P(\mathbf{G})$ factorizes as a simple product over pairs of vertices. This has the nice consequence that the expression for $\langle X\rangle$ is generally only as complicated as that for $X$. In other words, after the preliminary maximum-likelihood estimation of the parameters $\{\theta_a\}$, in this method the time required to obtain the exact expectation value of an $O(N^\tau)$ property across the entire randomized graph ensemble is the same as that required to measure the same property on the original real network, i.e. still $O(N^\tau)$. Therefore, as compared to the local rewiring algorithm, which requires a time $O(M\cdot N^\tau)$, the maximum-likelihood method is $O(M)$ times faster, for arbitrarily large $M$. 

\begin{table*}[]
\centering
\begin{tabular}{|c|c|}
\hline
\mbox{\textbf{Empirical undirected properties}} & \mbox{\textbf{Expected undirected properties}}\\
\hline
$a_{ij}$ & $\langle a_{ij}\rangle=p_{ij}=\frac{x_{i}x_{j}}{1+x_i x_j}$\\
\hline
$k_{i}=\sum_{j\ne i}a_{ij}$ & $\langle k_{i}\rangle=\sum_{j\ne i}p_{ij}$\\
\hline
$k_{i}^{nn}=\frac{\sum_{j\ne i}a_{ij}k_{j}}{k_{i}}$ & $\langle k_{i}^{nn}\rangle=\frac{\sum_{j\ne i}p_{ij}k_{j}}{\langle k_{i}\rangle}$\\
\hline
$c_{i}=\frac{\sum_{j\ne i}\sum_{k\ne i,j}a_{ij}a_{jk}a_{ki}}{\sum_{j\ne i}\sum_{k\ne i,j}a_{ij}a_{ik}}$ & $\langle c_{i}\rangle=\frac{\sum_{j\ne i}\sum_{k\ne i,j}p_{ij}p_{jk}p_{ki}}{\sum_{j\ne i}\sum_{k\ne i,j}p_{ij}p_{ik}}$\\
\hline
\mbox{\textbf{Empirical directed properties}} & \mbox{\textbf{Expected directed properties}}\\
\hline
$a_{ij}$ & $\langle a_{ij}\rangle=p_{ij}=\frac{x_{i}y_{j}}{1+x_{i}y_{j}}$\\
\hline
$k^{in}_{i}=\sum_{j\ne i}a_{ji}$ & $\langle k^{in}_{i}\rangle=\sum_{j\ne i}p_{ji}$\\
\hline
$k^{out}_{i}=\sum_{j\ne i}a_{ij}$ & $\langle k^{out}_{i}\rangle=\sum_{j\ne i}p_{ij}$\\
\hline
${k}_{i}^{tot}={k}_{i}^{in}+{k}_{i}^{out}$ & $\langle {k}_{i}^{tot}\rangle=\langle{k}_{i}^{in}\rangle
+\langle{k}_{i}^{out}\rangle={k}_{i}^{tot}$\\
\hline
$k^{\leftrightarrow}_{i}=\sum_{j\ne i}a_{ij}a_{ji}$ & $\langle k^{\leftrightarrow}_{i}\rangle=\sum_{j\ne i}p_{ij}p_{ji}$\\
\hline
$k_{i}^{in/in}=\frac{\sum_{j\ne i}a_{ji}k_{j}^{in}}{k_{i}^{in}}$ & $\langle k_{i}^{in/in}\rangle=\frac{\sum_{j\ne i}p_{ji}k_{j}^{in}}{\langle k_{i}^{in}\rangle}$\\
\hline
$k_{i}^{in/out}=\frac{\sum_{j\ne i}a_{ji}k_{j}^{out}}{k_{i}^{in}}$ & $\langle k_{i}^{in/out}\rangle=\frac{\sum_{j\ne i}p_{ji}k_{j}^{out}}{\langle k_{i}^{in}\rangle}$\\
\hline
$k_{i}^{out/in}=\frac{\sum_{j\ne i}a_{ij}k_{j}^{in}}{k_{i}^{out}}$ & $\langle k_{i}^{out/in}\rangle=\frac{\sum_{j\ne i}p_{ij}k_{j}^{in}}{\langle k_{i}^{out}\rangle}$\\
\hline
$k_{i}^{out/out}=\frac{\sum_{j\ne i}a_{ij}k_{j}^{out}}{k_{i}^{out}}$ & $\langle k_{i}^{out/out}\rangle=\frac{\sum_{j\ne i}p_{ij}k_{j}^{out}}{\langle k_{i}^{out}\rangle}$\\
\hline
$k_{i}^{tot/tot}=\frac{\sum_{j\ne i}(a_{ij}+a_{ji})k_{j}^{tot}}{k_{i}^{tot}}$ & $\langle k_{i}^{tot/tot}\rangle=\frac{\sum_{j\ne i}(p_{ij}+p_{ji})k_{j}^{tot}}{\langle k_{i}^{tot}\rangle}$\\
\hline
$c_{i}^{in}=\frac{\sum_{j\ne i}\sum_{k\ne i,j}a_{jk}a_{ji}a_{ki}}{k_{i}^{in}(k_{i}^{in}-1)}$ & $\langle c_{i}^{in}\rangle=\frac{\sum_{j\ne i}\sum_{k\ne i,j}p_{jk}p_{ji}p_{ki}}{\sum_{j\ne i}\sum_{k\ne i,j}p_{ji}p_{ki}}$\\
\hline
$c_{i}^{out}=\frac{\sum_{j\ne i}\sum_{k\ne i,j}a_{ik}a_{ij}a_{jk}}{k_{i}^{out}(k_{i}^{out}-1)}$ & $\langle c_{i}^{out}\rangle=\frac{\sum_{j\ne i}\sum_{k\ne i,j}p_{ik}p_{ij}p_{jk}}{\sum_{j\ne i}\sum_{k\ne i,j}p_{ij}p_{ik}}$\\
\hline
$c_{i}^{cyc}=\frac{\sum_{j\ne i}\sum_{k\ne i,j}a_{ij}a_{jk}a_{ki}}{k_{i}^{in}k_{i}^{out}-k_{i}^{\leftrightarrow}}$ & $\langle c_{i}^{cyc}\rangle=\frac{\sum_{j\ne i}\sum_{k\ne i,j}p_{ij}p_{jk}p_{ki}}{\langle k_{i}^{in}\rangle \langle k_{i}^{out}\rangle-\sum_{j\ne i}p_{ij}p_{ji}}$\\
\hline
$c_{i}^{mid}=\frac{\sum_{j\ne i}\sum_{k\ne i,j}a_{ik}a_{jk}a_{ji}}{k_{i}^{in}k_{i}^{out}-k_{i}^{\leftrightarrow}}$ & $\langle c_{i}^{mid}\rangle=\frac{\sum_{j\ne i}\sum_{k\ne i,j}p_{ik}p_{jk}p_{ji}}{k_{i}^{in}k_{i}^{out}-\sum_{j\ne i}p_{ij}p_{ji}}$\\
\hline
$c_{i}^{tot}=\frac{\sum_{j\ne i}\sum_{k\ne i,j}(a_{ij}+a_{ji})(a_{jk}+a_{kj})(a_{ki}+a_{ik})}{2\big[k_{i}^{tot}(k_{i}^{tot}-1)-2 k_{i}^{\leftrightarrow}\big]}$ & $\langle c_{i}^{tot}\rangle=\frac{\sum_{j\ne i}\sum_{k\ne i,j}(p_{ij}+p_{ji})(p_{jk}+p_{kj})(p_{ki}+p_{ik})}{2\big[\sum_{j\ne i}\sum_{k\ne i,j}(p_{ji}p_{ki}+p_{ij}p_{ik})+2(k_{i}^{in}k_{i}^{out})-2\sum_{j\ne i}p_{ij}p_{ji}\big]}$\\
\hline
\end{tabular}
\caption{Expressions for the empirical and expected properties in the binary (undirected and directed) representations of the network.\label{tab_b}}
\end{table*}

\section{Binary undirected properties\label{app_bun}}
In the binary undirected case, each graph $\mathbf{G}$ is completely specified by its (symmetric) Boolean adjacency matrix $\mathbf{A}$.
The randomization method described above proceeds by 
\begin{enumerate}
\item \textit{specifying the degree sequence as the constraint:} $\{C_a\}=\{k_i\}$. The Hamiltonian therefore reads
\begin{equation}
H(\mathbf{A})=\sum_i\theta_i k_i(\mathbf{A})=\sum_i\sum_{j<i}(\theta_i+\theta_j)a_{ij}
\end{equation}
and one can show \cite{newman_expo} that this allows to write the graph probability as
\begin{equation}
P(\mathbf{A})=\prod_{i}\prod_{j<i} p_{ij}^{a_{ij}}(1-p_{ij})^{1-a_{ij}}
\end{equation}
where
\begin{equation}
p_{ij}=\frac{x_i x_j}{1+x_i x_j}
\label{eq_bun_pij}
\end{equation}
(with $x_i\equiv e^{-\theta_i}$) is the probability that a link exists between vertices $i$ and $j$  in the maximum-entropy ensemble of binary undirected graphs, subject to specifying a given degree sequence as the constraint;

\item \textit{solving the maximum-likelihood equations,} by setting the parameters $\{x_i\}$ to the values
that maximize the likelihood $P(\mathbf{A}^*)$ to obtain the real network ${A}^*$ \cite{myrandomization,mylikelihood}. These values can be found as the solution to the following set of $N$ coupled nonlinear equations \cite{mylikelihood}:
\begin{equation}
\langle k_i\rangle=\sum_{j\ne i}\frac{x_i x_j }{1+x_i x_j}
=k_i(\mathbf{A}^*)\qquad\forall i
\label{eq_meank}
\end{equation}
where $\{k_i(\mathbf{A}^*)\}$ is the empirical degree sequence of the real network $\mathbf{A}^*$. For a detailed analysis about solving such system see \cite{fienberg} (for a discussion about the existence of solutions) and \cite{myrandomization} (for a discussion about the convergence of the algorithm). In principle, dimensionality and memory problems can arise when $N$ is too large (luckily, this is not the case of the ITN considered here). In such a case, the system can be re-written with a lower number of equation to solve. In fact, the hidden variables of the vertices with the same degree have the same value. So, one can straightforwardly solve the system only for them \cite{mylikelihood};

\item \textit{computing the probability coefficients $p_{ij}$}, by inserting the values $\{x_i\}$ into eq.~(\ref{eq_bun_pij}) which allows to easily compute the expectation value $\langle X\rangle$ of any topological property $X$ analytically, without generating the randomized networks explicitly \cite{myrandomization}. With this choice, eq.~(\ref{eq_bun_pij}) yields the exact value of the connection probability in the ensemble of randomized networks with the same average degree sequence as the empirical one. 
Note that $p_{ij}$ is the probability of a link between vertex $i$ and vertex $j$ in the grandcanonical ensemble (which is directly obtained analytically), and not the frequency of such a link in the corresponding microcanonical ensemble (which would require the explicit generation of artificially rewired networks). In ref.\cite{myrandomization}, it was shown that the microcanonical frequency converges to $p_{ij}$ asymptotically as the number of randomization steps in the microcanonical algorithm increases.
Equation (\ref{eq_meank}) shows that, by construction, the degrees of all vertices are special local quantities whose expected and empirical values are exactly equal: $\langle k_i\rangle=k_i$. It follows that the $p_{ij}$ coefficients can be calculated by using any of the networks in the corresponding microcanonical ensemble with constrained degree sequence: the expected values of the high-order properties will be the same;

\item \textit{computing the expectation values of higher-order topological properties}, as in Table \ref{tab_b}. The expressions are derived exploiting the fact that $\langle a_{ij}\rangle=p_{ij}$, and that different pairs of vertices are statistically independent, which implies $\langle a_{ij}a_{kl}\rangle=p_{ij}p_{kl}$ if $(i-j)$ and $(k-l)$ are distinct pairs of vertices, whereas $\langle a_{ij}a_{kl}\rangle=\langle a_{ij}^2\rangle=\langle a_{ij}\rangle=p_{ij}$ if $(i-j)$ and $(k-l)$ are the same pair of vertices. Also, the expected value of the ratio of two quantities is approximated with the ratio of the expected values: $\langle n/d\rangle\approx\langle n\rangle/\langle d\rangle$.
\end{enumerate}

\section{Binary directed properties\label{app_bdn}}
In the binary directed case, the above results can be generalized as follows.
Each graph $\mathbf{G}$ is completely specified by its Boolean adjacency matrix $\mathbf{A}$, which now is in general not symmetric. The maximum-likelihood randomization method \cite{myrandomization} proceeds in this case by
\begin{enumerate}
\item \textit{specifying both the in-degree and the out-degree sequences as the constraints:} $\{C_a\}=\{k^{in}_i,k^{out}_i\}$. The Hamiltonian takes the form
\begin{eqnarray}
H(\mathbf{A})&=&\sum_i\left[\theta^{in}_i k^{in}_i(\mathbf{A})+\theta^{out}_i k^{out}_i(\mathbf{A})\right]\nonumber\\
&=&\sum_i\sum_{j\ne i}(\theta^{in}_i+\theta^{out}_j)a_{ij}
\end{eqnarray}
The above choice leads to the graph probability \cite{myrandomization}
\begin{equation}
P(\mathbf{A})=\prod_{i}\prod_{j\ne i} p_{ij}^{a_{ij}}(1-p_{ij})^{1-a_{ij}}
\end{equation}
where
\begin{equation}
p_{ij}=\frac{x_i y_j}{1+x_i y_j}
\label{eq_bdn_pij}
\end{equation}
(with $x_i\equiv e^{-\theta^{out}_i}$ and $y_i\equiv e^{-\theta^{in}_i}$) is the probability that a link exists from vertex $i$ to vertex $j$  in the maximum-entropy ensemble of binary directed graphs with specified in- and out-degree sequences;

\item \textit{solving the maximum-likelihood equations,} by setting the parameters $\{x_i\}$ and $\{y_i\}$ to the values
that maximize the likelihood $P(\mathbf{A}^*)$ to obtain the real network ${A}^*$ \cite{myrandomization,mylikelihood}. These values can be found as the solution to the following set of $2N$ coupled nonlinear equations \cite{mylikelihood}:
\begin{eqnarray}
\langle k^{out}_i\rangle&=&\sum_{j\ne i}\frac{x_i y_j }{1+x_i y_j}
=k^{out}_i(\mathbf{A}^*)\qquad\forall i\label{eq_meankout}
\\
\langle k^{in}_i\rangle&=&\sum_{j\ne i}\frac{x_j y_i }{1+x_j y_i}
=k^{in}_i(\mathbf{A}^*)\qquad\forall i
\label{eq_meankin}
\end{eqnarray}
where $\{k^{in}_i(\mathbf{A}^*)\}$ and $\{k^{out}_i(\mathbf{A}^*)\}$ are the empirical degree sequences of the real  network $\mathbf{A}^*$. Again, for a detailed analysis about solving such system see \cite{fienberg,myrandomization};

\item \textit{computing the probability coefficients $p_{ij}$}, by inserting the values $\{x_i\}$ and $\{y_i\}$ into Eq.~(\ref{eq_bdn_pij}) which allows to easily compute the expectation value $\langle X\rangle$ of any topological property $X$ analytically, without generating the randomized networks explicitly \cite{myrandomization}. So Eq.~(\ref{eq_bdn_pij}) yields the exact value of the connection probability in the ensemble of randomized directed graphs with the same average degree sequences as the empirical ones and Eqs.~(\ref{eq_meankout})-(\ref{eq_meankin}) show that, by construction, the in-degrees and out-degrees of all vertices are special local quantities whose expected and empirical values are exactly equal: $\langle k^{in}_i\rangle=k^{in}_i$ and $\langle k^{out}_i\rangle=k^{out}_i$. It follows that the $p_{ij}$ coefficients can be calculated by using any of the networks in the corresponding microcanonical ensemble with constrained in-degree and out-degree sequences: the expected values of the high-order properties will be the same;

\item \textit{computing the expectation values of the higher-order topological properties}, as in Table \ref{tab_b}, by using the same prescription as in the undirected case plus the additional care that now $(i-j)$ and $(j-i)$ are different (and statistically independent) directed pairs of vertices. Therefore $\langle a_{ij}a_{ji}\rangle=p_{ij}p_{ji}$.
\end{enumerate}

\bibliographystyle{apsrev}
\bibliography{rewiring_part1}

\end{document}